\documentclass[iop,apj,twocolappendix]{emulateapj}
\pdfoutput=1 
\usepackage{amsmath,amssymb,amstext}

\usepackage[breaklinks,colorlinks,citecolor=blue,linkcolor=magenta]{hyperref} 

\usepackage[all]{hypcap} 

\usepackage{aas_macros}
\usepackage{natbib}

\usepackage[table]{xcolor}

\shorttitle{CR Acceleration at Galactic Wind Termination Shocks}
\shortauthors{Bustard et al.}

\begin{document}

\title{Cosmic Ray Acceleration by a Versatile Family of Galactic Wind Termination Shocks}

\author{
Chad Bustard\altaffilmark{1}, 
Ellen G. Zweibel\altaffilmark{1,2}, and
Cory Cotter\altaffilmark{2}
}
\altaffiltext{1}{Physics Department, University of Wisconsin-Madison, 1150 University Avenue, Madison, WI 53706; bustard@wisc.edu}
\altaffiltext{2}{Department of Astronomy, University of Wisconsin-Madison, 2535 Sterling Hall, 475 N. Charter Street, Madison, WI 53706}

\begin{abstract}
There are two distinct breaks in the cosmic ray (CR) spectrum: the so-called ``knee" around $3 \times 10^{15}$ eV and the so-called ``ankle" around $10^{18}$ eV. Diffusive shock acceleration (DSA) at supernova remnant (SNR) shock fronts is thought to accelerate galactic CRs to energies below the knee, while an extragalactic origin is presumed for CRs with energies beyond the ankle. CRs with energies between $3 \times 10^{15}$ and $10^{18}$ eV, which we dub the ``shin," have an unknown origin. It has been proposed that DSA at galactic wind termination shocks, rather than at SNR shocks, may accelerate CRs to these energies. This paper uses the galactic wind model of \cite{2016ApJ...819...29B} to analyze whether galactic wind termination shocks may accelerate CRs to shin energies within a reasonable acceleration time and whether such CRs can subsequently diffuse back to the galaxy. We argue for acceleration times on the order of 100 Myrs rather than a few billion years, as assumed in some previous works, and we discuss prospects for magnetic field amplification at the shock front. Ultimately, we generously assume that the magnetic field is amplified to equipartition. This formalism allows us to obtain analytic formulae, applicable to any wind model, for CR acceleration. Even with generous assumptions, we find that very high wind velocities are required to set up the necessary conditions for acceleration beyond $10^{17}$ eV. We also estimate the luminosities of CRs accelerated by outflow termination shocks, including estimates for the Milky Way wind. 

\end{abstract}



\section{Introduction}
Cosmic rays (CRs) in the galaxy have an energy density in the interstellar medium (ISM) roughly in equipartition with magnetic energy and thermal energy, making them a fundamental component of our galaxy. Amongst other known impacts of CRs, they are important for heating and ionizing the ISM and for aiding in launching galactic outflows that regulate star formation(\cite{2013PhPl...20e5501Z}, \cite{2015ARA&A..53..199G}). Despite their importance to the dynamics of the ISM and their discovery more than a hundred years ago, the origin of CRs at various energies is still unknown. The main reason for this is that the distribution of CRs is highly isotropic due to resonant scattering off small-scale magnetic fluctuations, making it very difficult to pinpoint the sources of CRs observed on Earth. At CR energies less than $\approx 100$ GeV, these fluctuations can be self-excited by a cosmic ray streaming anisotropy (\cite{1968ApJ...152..987W}, \cite{1969ApJ...156..445K}, \cite{2003ApJ...587..625Z}, \cite{2010ApJ...709.1412Z}, \cite{1971ApJ...170..265S}), whereas at higher energies, extrinsic turbulence is the primary scatterer (\cite{2013PhPl...20e5501Z}). 

Despite this incredible complication, much has been discovered and subsequently accepted about CRs. In particular, the main source for galactic cosmic rays is most likely diffusive shock acceleration (DSA) at supernova remnant (SNR) shock fronts (\cite{1977ICRC...11..132A}, \cite{1978MNRAS.182..147B}, \cite{1978ApJ...221L..29B}). This process likely explains the origin of CRs at energies below $3 \times 10^{15}$ eV, which is called the ``knee" of the CR spectrum. Around this energy, the CR spectrum steepens. A second break in the spectrum occurs around $10^{18}$ eV, at the ``ankle," where the spectrum flattens. Above this energy, all cosmic rays are believed to have an extragalactic origin, because these particles will have gyroradii $r_{g} = \frac{eB}{\gamma mc}$ greater than the size of the Milky Way and therefore cannot be confined within the Galaxy. One can imagine CRs accelerated to these high energies within the galaxy, but the energy requirements are very large (\cite{2011ARA&A..49..119K}). 

The origin of CRs with energy between the knee and the ankle, which we will refer to as the ``shin" of the CR spectrum, is still unknown. Many possible solutions to this problem have been proposed. One natural solution is that most SNRs accelerate CRs only up to the knee, but smaller and smaller subsets of SNRs can accelerate particles to higher energies possibly on the order of $10^{17}$ eV (e.g. \cite{2014NuPhS.256..197P}). For example, \cite{2009A&A...499..191T} suggests that the shock of SN 1993J may have been able to accelerate CR protons to $2-3 \times 10^{16}$ eV within the first few days of the burst. 

Another interesting idea is that CRs are accelerated in superbubbles (SBs), which have the unique quality that their energy source is combined from the supernovae whose outbursts coalesce into SBs as well as from stellar winds. It is well known that SBs suffer from severe energy losses, which may be the result of very efficient particle acceleration that exhausts the SB shock energy (\cite{2008ApJ...677L..21B}). 

The natural extension of these SNR and SB shock front mechanisms to larger scales is galactic wind termination shocks. Galactic winds from supernova rich starburst regions and active galactic nuclei transfer mass and energy away from regions of wind development and enrich the intergalactic medium (IGM) with metals, eventually suppressing star formation in the galaxies (\cite{2005ARA&A..43..769V}, \cite{2015Natur.523..169E}). In addition to the feedback galactic outflows can provide for galaxy formation, DSA at galactic wind termination shocks has been proposed as a possible mechanism for accelerating high energy CRs between the knee and the ankle (\cite{1985ApJ...290L...1J}, \cite{1987ApJ...312..170J}; \cite{2004A&A...417..807V}). Specifically, this mechanism, at least on the surface, satisfies the two main ingredients required for efficient CR acceleration: good confinement and long acceleration time. 

Given the considerable amount of recent work on galactic outflows, we are in a good position to revisit the idea of galactic wind termination shock acceleration of CRs and test that theory's validity from a modern perspective. This paper attempts to do that and is organized as follows: Section \ref{windModel} describes the model (\cite{2016ApJ...819...29B}; hereafter BuZD16) we use to generate wind solutions and how that leads to a rough estimate of the termination shock radius. In Section \ref{taccel}, we discuss CR confinement and elaborate on our choice for the acceleration time based on the observed and simulated lifetimes of galactic outflows. A short analysis of magnetic field amplification mechanisms, which may be very important for CR acceleration, is given in Section \ref{amplify}.  Ultimately, when we estimate maximum CR acceleration rates using our model, we choose to simply allow the magnetic field to reach equipartition, thereby providing a generous upper bound on the rate of CR acceleration. In Section \ref{shockConditions}, we present estimates of the accelerated CR energy and define a CR Reynolds number, which gives an idea of whether accelerated particles can diffuse back to the galaxy in opposition to advection with the flow. This section is independent of wind model and can be used in combination with any wind model to connect fundamental wind parameters to CR acceleration. Section \ref{galaxyConditions} makes this connection with our model to generate outflows conducive to shock acceleration. Section \ref{lums} then provides estimates of the CR luminosity due to termination shocks. A discussion of the results and alternative prospects for CR acceleration is given in Section \ref{conclusions}. 

\section{Wind Model}
\label{windModel}
The galactic wind model used here is a re-working of the spherically symmetric \cite{1985Natur.317...44C} model (hereafter referred to as the CC model), including non-uniform mass and energy source distributions, a gravitational potential from an extended mass distribution, and radiative losses (see BuZD16 for a full description of the wind model). Recent improvements include the addition of a Hernquist potential (\cite{1990ApJ...356..359H}) to model the dark matter halo, and most important for this research, the interaction of the wind fluid with the IGM, resulting in a termination shock. 

The gravitational field used in BuZD16 is that of a constant density sphere of radius $R$ and mass $M$.   Adding the halo, the  gravitational field is: 
\begin{align}
   \frac{d\Phi}{dr} = \frac{G M_{h}}{(r+a)^{2}} + \frac{GMr}{R^{3}} \qquad & \text{for } r < R \\
    \frac{d\Phi}{dr} = \frac{G M_{h}}{(r+a)^{2}} + \frac{GM}{r^{2}} \qquad & \text{for } r > R
\end{align}
where  $M_{h}$ is the halo mass, and $a$ is a scale length determined by a semilog fit to the points $(10^{11} M_{\odot}, 6 \rm kpc)$ and $(10^{13} M_{\odot}, 25 \rm kpc)$. In Section \ref{galaxyConditions}, we show that the halo profile has minimal impact on the wind dynamics. For the rest of this paper, we assume $M_{h} = 0$ and use the default values $R = 200$pc,
$M=10^9M_{\odot}$. 

Beyond the radius R, mass and energy injection into the wind from supernovae is assumed to be negligible, and the quadratic mass injection profile we adopt reflects this: 
\begin{align}
q(r) = q_{0}(1-\frac{r^{2}}{R^{2}}) \qquad & \text{for } r < R \\
q(r) = 0 \qquad & \text{for } r > R \\
\end{align}
Here q(r) is the mass per unit time per volume injected into the wind, $q_{0}$ is q(r) evaluated at the galactic center, and we assume the energy loading follows the same profile. 

We parameterize the mass-loading by 
\begin{equation}
\dot{M} = \beta  \text{SFR} (\text{M}_{\odot}\text{/yr})
\end{equation}
Then our mass-loading per volume factor, $q_{0}$, is calculated such that $\dot{M} = \int_{0}^{R} q_{0}(1-r^{2}/R^{2}) dV$.
\begin{equation}
q_{0} = \frac{\beta \text{SFR}}{\frac{8}{15}\pi R^{3}} = 1.60 \times 10^{-37} \beta \text{SFR}
\end{equation}
fixing $R = 200$ pc for each galaxy, regardless of the galaxy's mass and SFR. 

The BuZD16 model also allows energy addition, which we model by a heating term in the First Law of Thermodynamics 
\begin{equation}
\dot{E} = \alpha \text{SFR},
\end{equation}
and also includes radiative cooling. For all the models discussed in this paper, we assume $\alpha = 0$, but radiative energy loss can be quite important for the higher $\beta$
flows. 

Our implementation of radiative cooling assumes collisional ionization equilibrium (CIE). Preliminary tests of non-equilibrium cooling in our model suggest that ionization fractions may be very different than those assumed here for CIE; however, our estimated non-equilibrium cooling curves for highly radiating winds are very similar to the optically thin equilibrium cooling curve assumed here. Therefore, although non-equilibrium ionization may have drastic effects on the observational signatures of our winds, the wind solutions we obtain will likely behave qualitatively the same as the equilibrium solutions shown in this paper. This is mostly consistent with the more detailed analysis of non-equilibrium cooling in outflows done by \cite{2016MNRAS.460.2157O}. Further modifications to our model, such as increasing the wind metallicity above solar values, are left to future work. 

\subsection{Termination Shock}
When the total wind pressure $\rho V^{2} + 3/2 n k_{B} T + B^{2}/8 \pi  \approx P_{\rm IGM}$, the pressure of the IGM, the wind forms a shock. For accuracy, we have included the magnetic pressure in this condition, but magnetic fields are not included in our simple wind models. The pressure of the IGM is not well-known, but equating the wind pressure with a reasonable value of $P_{\rm IGM} = 10^{-14} - 10^{-15} \text{ergs}$  $\rm cm^{-3}$ (\cite{1980ApJS...42...41S}, \cite{2002ApJ...573..157N}) gives us the distance to the termination shock, hereafter referred to as $R_{\rm shock}$. Assuming the magnetic field and thermal contributions to the total pressure is negligible, which is a fair assumption given the low magnetic field strengths and temperatures expected in the IGM, the shock radius is roughly given by

\begin{equation}
\label{shockradius}
R_{\rm shock} = \left(\frac{\dot{M} V}{4\pi P_{\rm IGM}}\right)^{1/2}\approx 58\frac{\dot M_{\odot}^{1/2}T_{07}^{0.263}}{P_{-14}^{1/2}}{\rm kpc},
\end{equation}
where in the first equality the wind velocity, $V = V_{\rm shock}$, is obtained from the wind model and in the second equality we have used the scaling law from Figure \ref{T0_vs_ShockVel_vs_Beta} relating asymptotic velocity and central temperature $T_0$, and expressed $\dot M$ in $M_{\odot}/yr$, $T_0$ in $10^7K$, and $P_{\rm IGM}$ in $10^{-14}$ dyne cm$^{-2}$. 
 
To estimate if these shocks will be radiative, we compare the cooling time $t_{\rm cool} = 3k_{B}T/n\Lambda$ of the wind to the dynamical time $t_{\rm dyn} \approx R_{\rm shock}/V_{\rm shock}$. 
Using Equation (\ref{shockradius}) and the Rankine-Hugoniot conditions for a strong shock in a $\gamma = 5/3$ gas, we find
\begin{equation}\label{tctd}
\frac{t_{\rm cool}}{t_{\rm dyn}}=\frac{9}{128}\frac{(4\pi)^{1/2}m^2V_{\rm shock}^{4.5}}{P_{\rm IGM}^{1/2}\dot M^{1/2}\Lambda}\approx 220\frac{T_{07}^{2.25}}{P_{-14}^{1/2}\dot M_{\odot}^{1/2}\Lambda_{-22}},
\end{equation} 
which comfortably exceeds unity for most of the wind models in our parameter space. Therefore, we ignore post-shock radiative cooling. 

\section{Confinement and Acceleration Time}
\label{taccel}

The radius at which a galactic wind termination shock occurs is typically at least a hundred kpc, much greater than the radius of the Galaxy; hence CRs accelerated by DSA at this termination shock need only have gyroradii comparable to or less than a hundred kpc to be confined to the termination shock, and therefore can be accelerated to higher energies than CRs accelerated by SNRs.  

The maximum energy of confined protons at a shock distance of $R_{\rm shock}$ is
\begin{equation}
E_{\rm GeV} = \frac{10^{-2}B_{\mu G} R_{\rm shock}}{3 \times 10^{10}}
\end{equation}
where $E_{\rm GeV}$ is the CR energy in GeV, $B_{\mu G}$ is the magnetic field strength in microgauss, and $R_{\rm shock}$ is the radius of the shock in cm. Let's make a generous estimate of this CR energy assuming the magnetic field strength in a galaxy at a radius of $200$ pc is 300 $\mu$G. This galactic magnetic field value is consistent with models of M82 developed by \cite{2016MNRAS.457L..29Y}. Assuming the radial magnetic field component drops off as $1/r^{2}$, while the azimuthal component will drop off as $1/r$, then at $\approx$ 200 kpc, the azimuthal component should dominate the radial component. Then, at 200 kpc, $B \approx 0.3$ $\mu$G. Plugging this into the equation for accelerated proton energy, the maximum energy of a confined CR at a distance of 200 kpc with $B \approx 0.3$ $\mu$G will be about $10^{19}$ eV. If we instead use a shock distance of $R_{\rm shock} = 20$ kpc, we can say that $10^{17}$ eV protons can be confined. 

Confinement, then, may be an issue for the highest energy CRs closer to the ankle if they are accelerated close to the galactic disk; however, presumably the magnetic field strength is also greater near the galactic disk. Overall, we see that confinement is not such a strict requirement for acceleration. 

The acceleration time, however, is much more limiting. \cite{1983A&A...125..249L} derived the maximum rate of particle acceleration by a strong parallel shock, which can be written in the following form (\cite{2003ApJ...587..625Z}):
\begin{equation}
\label{Legage}
\frac{dE}{dt} = 1.5 \times 10^{-18} Z B V_{\rm shock}^{2} \rm GeV/s
\end{equation}
where B is in Gauss, and the shock velocity is in cm/s. If we assume the particles are accelerated at this maximum rate for an acceleration time $t_{\rm acc}$, then 

\begin{equation}
\label{EMax}
E_{\rm max} = \frac{dE}{dt} t_{\rm acc}
\end{equation}
Equations (\ref{Legage}) and (\ref{EMax}) gives us two important clues as to how we may increase the maximum energy of DSA CRs. The first way is to increase the magnetic field. In Section \ref{amplify} we will discuss field amplification mechanisms which could operate near the shock front, possibly up to equipartition with the shock itself. Here we note that in the
absence of field amplification, DSA is probably rather slow. For the 0.3 $\mu G$ field introduced above in discussing confinement and for $V_{\rm shock} = 1000$ km/s, Equation (\ref{Legage}) predicts an energy gain of 1.4 PeV in 100 Myr  - and this assumes a rather large magnetic field at the base of the wind. Therefore, it seems unlikely that DSA at galactic wind termination shocks can
produce cosmic rays at the shin without some degree of magnetic field amplification. 

The second way to increase the maximum CR energy is to accelerate them for long periods of time. As an example, we can estimate the maximum energy of particles accelerated by a plane parallel, galactic wind termination shock with wind velocity $V = V_{\rm shock} = 1000$ km/s and $B = 0.3$ $\mu$G. Using Equation (\ref{Legage}) to get dE/dt, 
\begin{equation}
E_{\rm max} = (4.5 \text{eV/s}) \times t_{\rm acc}
\end{equation}
We see that to accelerate particles to $10^{19}$ eV will be to accelerate them for $10^{19}$ seconds, or greater than the age of the universe. To accelerate particles to $10^{17}$ eV would require an acceleration time of a billion years, which is on the order of the age of the Milky Way. SNR lifetimes, however, are much shorter, ultimately leading to a CR energy cut-off slightly higher than $10^{15}$ eV in simulations even with the most effective magnetic field amplification. 

The lifetime of galaxies is inherently longer than the lifetimes of stars. Hence, it seems plausible that galactic winds and their termination shocks would be sustained for much longer times than their stellar counterparts, SNRs. Substantial progress in models and observations of galactic outflows has been made, however, since the notion of galactic winds as high energy CR accelerators was proposed by \cite{1985ApJ...290L...1J}. Importantly, there is a growing consensus that the lifetimes of individual galactic outflows is not comparable to the lifetimes of their host galaxies. Although supernovae driven outflows may be prevalent throughout the lifetimes of galaxies, they seem to be very bursty, typically accompanying intense but short episodes of star formation (\cite{2015MNRAS.454.2691M}, \cite{2015MNRAS.453.3499K}, \cite{2016arXiv160204856R}, \cite{2016MNRAS.462..994S}. 

Using the FIRE (Feedback in Realistic Environments) simulations (\cite{2014MNRAS.445..581H}), \cite{2015MNRAS.454.2691M} find that, especially for high redshift galaxies, star formation occurs in sharp bursts, with the galaxy spending a significant amount of time at near zero SFR. As a result, the galaxy outflow rate is highly variable with outflows following star formation bursts and then shutting off. At redshift $\approx 0.75$, star formation becomes more continuous, however, with weak and continuous outflows with decreased mass-loading factor compared to the bursty outflow stage. These weak continuous outflows, due to having a low mass-loading factor, are generally less susceptible to catastrophic momentum loss due to radiative cooling, and will have generally higher velocities than high mass-loaded flows. Combined with a long lifetime, such outflows could result in efficient particle acceleration; however, further simulations are required to determine the stability and lifetimes of termination shocks under such conditions. 

Strong outflows, whether in the bursty high redshift phase described in \cite{2015MNRAS.454.2691M} or in a continuous phase, may be limited simply by the mass supply of the galaxy. M82, as an example, contains about $10^{8} M_{\odot}$ of stellar mass and has a wind mass-loading factor of order 10 $M_{\odot}$/yr in the central starburst region (\cite{2003ApJ...599..193F}, \cite{2004ApJ...606..271G}), meaning that a wind with this strength could only last for approximately 0.1 Gyrs until it exhausts the internal mass supply of M82 (\cite{2013ApJ...768...53Y}). New high resolution images of M82 suggest an even higher outflow rate that could deplete the galaxy of molecular gas within 8 Myrs (\cite{2016arXiv160800974C}. Overall, this suggests that wind lifetimes are more realistically on the order of 100 Myrs, much shorter than the lifetimes of galaxies, which are on the order of Gyrs.

\section{Magnetic Field Amplification}
\label{amplify}

It has become clear in recent years that magnetic field amplification takes place in SNRs and may play a significant role in generating the turbulent scattering on either side of the shock that is necessary for DSA to occur. 
Magnetic field amplification has been observed in supernova remnants. In Cas A, equipartition arguments to explain the radio luminosity imply magnetic fields of a few mG, and further investigation of the nonthermal X-ray emission in this remnant suggest magnetic field strengths of $0.08 - 0.16$ mG in the rim (\cite{2003ApJ...584..758V}), far above what is expected in the shocked interstellar medium. 

The exact amplification mechanism is still unknown; however, a few theories have emerged: The Bell instability (\cite{2004MNRAS.353..550B}, \cite{2009MNRAS.392.1591A}), also known as the cosmic ray current driven instability, and turbulent amplification generated by density fluctuations (\cite{2007ApJ...663L..41G}) and enhanced by a CR pressure (\cite{2009ApJ...707.1541B}) in the clumpy medium near the shock. Before going into more detail, it is important to note whether these theories amplify the field upstream or downstream of the shock. The Bell instability requires positively charged CRs propagating in the magnetized region upstream of the shock; therefore, the amplification is \emph{upstream} of the shock. Also relying on propagating CRs, the turbulent amplification described in \cite{2009ApJ...707.1541B} acts \emph{upstream} of the shock, whereas the turbulent amplification described by \cite{2007ApJ...663L..41G} acts in the \emph{downstream} region. Further work describing amplification downstream can be found in \cite{2012ApJ...747...98G} and \cite{2016MNRAS.tmp.1426J}. 

\subsection{Bell Instability}
The Bell instability amplifies low-frequency, right circularly polarized waves with direction of propagation parallel to the magnetic field. Unlike the gyroresonant streaming instability (\cite{1969ApJ...156..445K}), it acts at wavelengths much less than the cosmic ray gyroradius. It has been proposed that this instability could possibly amplify the magnetic field strength by up to two orders of magnitude and aid in accelerating CRs from SNRs up to the knee (\cite{2005AIPC..801..337O}, \cite{2008MNRAS.386..509R}, \cite{2010ApJ...709.1412Z}).

In the cold plasma limit,  which should be a good approximation here due to the strong radiative and adiabatic cooling of the wind, the criterion to excite the Bell instability can be expressed as
\begin{equation}
\label{bellcriterion}
\frac{U_B}{U_{cr}} < \frac{v_D}{c},
\end{equation}
where $U_B$ and $U_{cr}$ are the magnetic field and cosmic ray energy densities, respectively, and 
$v_{D}$ is the drift velocity, which we can simply equate to the shock velocity $V_{\rm shock}$. For young SNR shock fronts, which have higher velocities than 
our estimates for wind termination shocks, the Alfv{\'e}n speed and drift velocity will be small and large, respectively, making the Bell instability criterion more difficult to satisfy for
termination shocks than for young SNRs.

If the instability is excited, the growth rate is generally fast. In the cold plasma limit, and for $v_D/v_A\gg 1$, the maximum growth rate is independent of  magnetic field strength, and
is given by 
\begin{equation}
\label{Bellgrowth}
\omega_{Bell} = \frac{\omega_{ci} n_{cr}}{2 n_{i}} \frac{v_{D}}{v_{A}}
\end{equation}
For wind parameters and large magnetic field strengths appropriate for M82 (\cite{2016MNRAS.457L..29Y}), the growth time should be very short. 

To estimate the saturated magnetic field strength, it is logical (though the situation is a little more complicated than this) to assume the magnetic field is maximal when the Bell instability criterion is no longer valid. This occurs when

\begin{equation}
\frac{B^{2}}{4\pi} \approx \frac{P_{\rm cr} V_{\rm shock}}{c} 
\end{equation}

If the CR pressure were roughly equal to the ram pressure of the flow $\rho V_{\rm shock}^{2}$, which is a generous upper limit, the magnetic field would be
\begin{equation}
\label{BellSaturation}
\frac{B^{2}}{4\pi} = (\frac{V_{\rm shock}}{c}) \rho V_{\rm shock}^{2}
\end{equation}
For wind velocities on the order of $10^{8}$ cm/s, this represents magnetic field amplification to only $\sim1\%$ of equipartition.

Therefore, although the Bell instability could grow within a very short time, the CR density must be quite high to excite the instability, and the resulting amplification is not very high. It needs to be noted, however, that saturation of the Bell instability is very uncertain. Among other things, the saturation depends on damping, the back reaction of the amplified field on the CR current (\cite{2009ApJ...694..626R}), and coupling between large and small spatial scales. Overall, this instability must rely on non-linear growth to achieve the large scale fields needed for particle acceleration because this instability saturates on scales near the particle gyroradius. 

A number of numerical simulations of the Bell instability have been carried out (\cite{2004MNRAS.353..550B}, \cite{2008ApJ...678..255Z}, \cite{2012ApJ...753....6R}, \cite{2014ApJ...788..107B}).  Direct numerical simulations of \cite{2012ApJ...753....6R} show magnetic field amplification beyond the rough estimate made above. They find that magnetic field amplification follows three stages: 1) the Bell instability is excited and amplifies the magnetic field; 2) linear growth continues among high-k (small spatial scale) modes, while mode coupling excites low-k (large scale) modes; 3) growth continues possibly due to an $\alpha^{2}$ dynamo. This turbulent growth could continue to amplify the magnetic field beyond the rough saturation limit in Equation (\ref{BellSaturation}) and, most importantly, amplify the magnetic field on large scales in addition to small scales. \cite{2014ApJ...788..107B} suggest that typical scales of Bell instability amplified fields in either the linear or non-linear growth regimes will be too small to affect the desired particle acceleration without taking into account back reaction to CRs, such as the CR pressure driven amplification discussed in the following section. 

\subsection{Turbulent Amplification}
Finally, let's discuss the possibility of turbulence, generated in a clumpy medium and enhanced by a CR pressure, amplifying the large-scale magnetic field near the termination shock. Whereas the Bell instability effectively relies on having non-magnetized CRs streaming through a strongly magnetized background plasma, CR pressure driving considers CRs that are strongly coupled to the background plasma through an effective pressure. This pressure accelerates regions of different density at different rates, and the resulting density fluctuations plus the CR pressure drives turbulence near the shock. This turbulence winds up the magnetic field, thereby amplifying it.

There are a few advantages to this process compared to amplification due to the non-resonant instability. One possible issue with the Bell instability stems from the scales on which it is excited. The instability acts on scales smaller than a particle gyroradius, meaning that some non-linear turbulent growth is then required to amplify the field at large scales. The CR pressure driven instability, however, acts on scales large compared to the gyroradius but small compared to the diffusion scale. Unlike many plasma instabilities (for example, the CR resonant streaming instability), this CR pressure driven instability is theoretically limited only by equipartition with the background medium.

This results in a stronger maximum magnetic field strength than that expected from the Bell instability (without turbulent growth considered). The results of \cite{2013MNRAS.436..294B} are conveniently scaled, and estimates of the magnetic field amplification and timescale on which this amplification occurs can be made for our termination shock set-up. This results in a code time unit of $\approx 200$ Myrs. Looking at Figure 4 from \cite{2013MNRAS.436..294B}, the magnetic field is amplified by a factor of 10 on average after $t = 0.5$ code units, or $\approx 100$ Myrs for our galactic wind model, with maximum amplification closer to a factor of 50. Similar levels of mean amplification have been achieved by \cite{2012MNRAS.427.2308D} and \cite{2016MNRAS.458.1645D}, amongst others.

\section{Particle Acceleration: Necessary Conditions at the Termination Shock}
\label{shockConditions}

Regardless of differences in growth times and saturation, which is not well-known, each of these mechanisms is ultimately limited approximately by equipartition. Therefore, \emph{for the sake of simplicity, we will optimistically assume the most generous magnetic field amplification, i.e. that the magnetic pressure is amplified to equal the ram pressure of the outflow}: 

\begin{equation} 
\label{equi}
\frac{B^{2}}{4\pi} = \rho V_{\rm shock}^{2}
\end{equation} 

The problem is then two-fold: First, given a certain intergalactic ambient pressure and a magnetic field amplified to equipartition, can we generate wind termination shocks at such a radius and velocity that the shocks could theoretically accelerate shin CRs within an acceleration time of $100$ Myrs? Second, to determine whether these CRs might be of galactic or extragalactic origin, we will ask how likely it is that those CRs can diffuse back towards Earth. Regardless of whether the CRs can diffuse back towards the galaxy, it is an interesting analysis just to answer the first question. As in our previous estimates, we will assume the CRs are constantly accelerated at the rate given by \cite{1983A&A...125..249L} throughout the acceleration time, and we will assume that the wind is in a steady state, which is an assumption of our wind model. We will show that certain wind solutions do give rise to possibly efficient CR acceleration at termination shocks. 

The results will be presented in two sections. First, we can draw a number of conclusions, independent of wind model, about the conditions necessary at the termination shock for efficient CR acceleration. These results are presented in the current section, Section \ref{shockConditions}, entitled ``Necessary Conditions at the Termination Shock." This section has particularly broad-ranging utility because interest in large scale shocks as CR accelerators is not limited to galactic wind shocks. For example, shocks created by the interaction of active galactic nuclei (AGN) driven outflows with the surrounding medium may also be efficient sites of CR acceleration (\cite{1999APh....11..347H}, \cite{2009ApJ...698L.138B}, \cite{2011ICRC....8...39P}, \cite{2016arXiv161107616W}). Because our formulation can be combined with any outflow model to estimate CR acceleration, the following machinery is not only applicable to galactic wind shocks but also to any large scale shock created by an outflow. In Section \ref{galaxyConditions}, we will specifically focus on galactic wind termination shocks and analyze CR acceleration with our thermally driven wind model, allowing us to derive a number of conclusions about the conditions necessary in the galaxy itself to produce favorable termination shock conditions.

First, let us assume an equipartition magnetic field and combine Equations (\ref{equi}) and (\ref{Legage}). At distances on the order of $100$ kpc at which the termination shocks typically occur, one can safely assume that the ram pressure dominates the magnetic and internal pressures of the wind; hence, at the shock, $\rho V_{\rm shock}^{2} \approx P_{\rm IGM}$. These steps give us $dE/dt$ as a function of just the shock velocity and the ambient pressure.

\begin{equation}
\label{dEdtEquipartition}
\frac{dE}{dt} = (1.5 \times 10^{-18} Z) \sqrt{4\pi P_{\rm IGM}} V_{\rm shock}^{2} \quad \text{GeV/s}
\end{equation}

\begin{figure}
\label{maxdEdt}
\centering
\includegraphics[scale = 0.43]{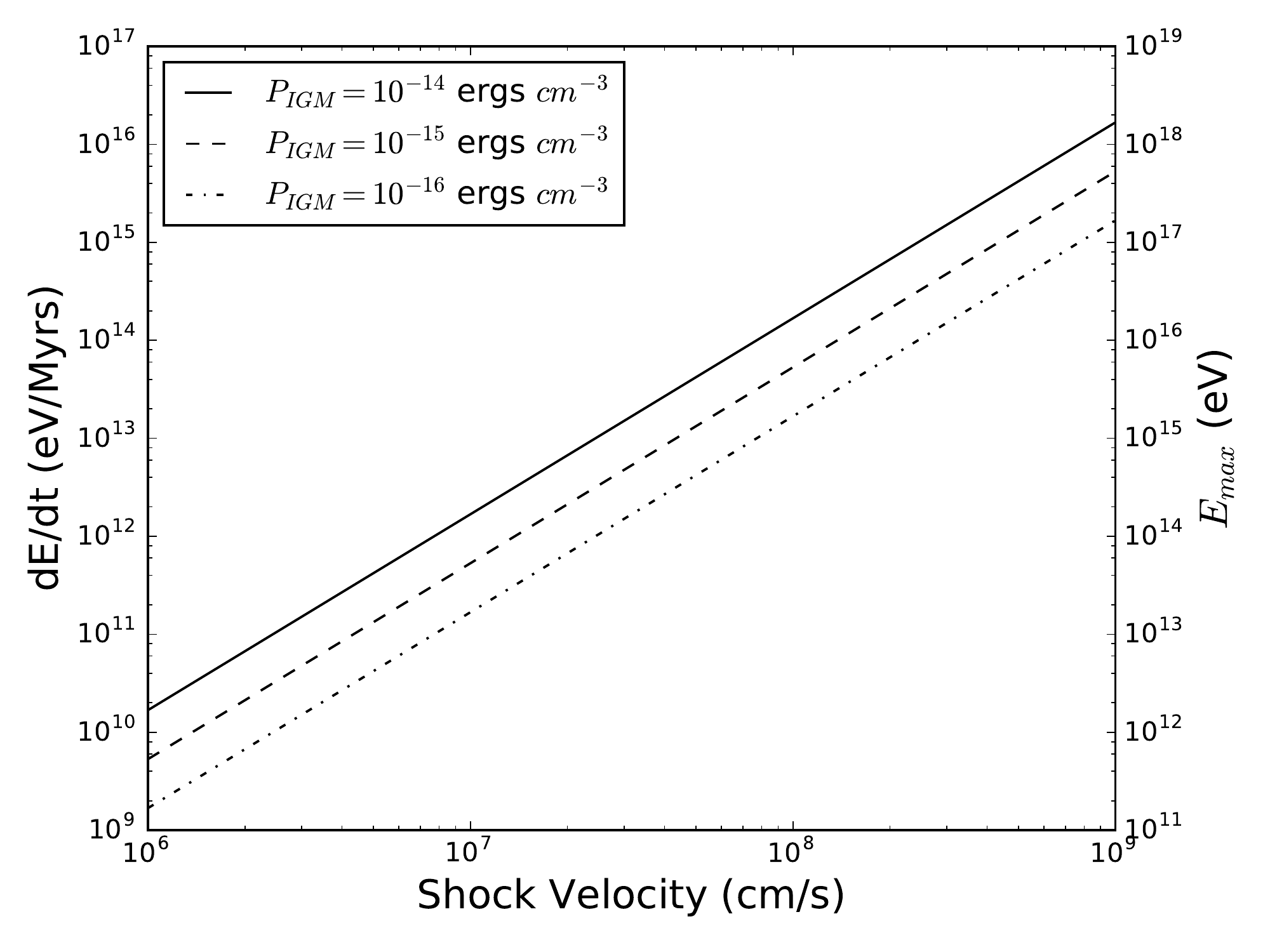}
\caption{Assuming the magnetic field is amplified to equipartition, and the ram pressure approximately equals the pressure of the surrounding IGM at the shock position, one can find the \cite{1983A&A...125..249L} dE/dt as a function of shock velocity for various $P_{\rm IGM}$.}
\end{figure}

A plot of $dE/dt$ in eV/Myrs as a function of shock velocity for various $P_{\rm IGM}$ is given in Figure \ref{maxdEdt}. Note that, given an acceleration time of 100 Myrs and $P_{\rm IGM} = 10^{-14}$ ergs $ \rm cm^{-3}$, a shock velocity of roughly 2000 km/s would be required to accelerate protons to $10^{17}$ eV. If we instead choose a lower ambient pressure, the required velocity is even higher for the same target CR energy. 

\begin{figure*}
\label{reynoldsStrong_vs_Rshock_vs_Ushock_varya}
\centering
\includegraphics[scale = 0.44]{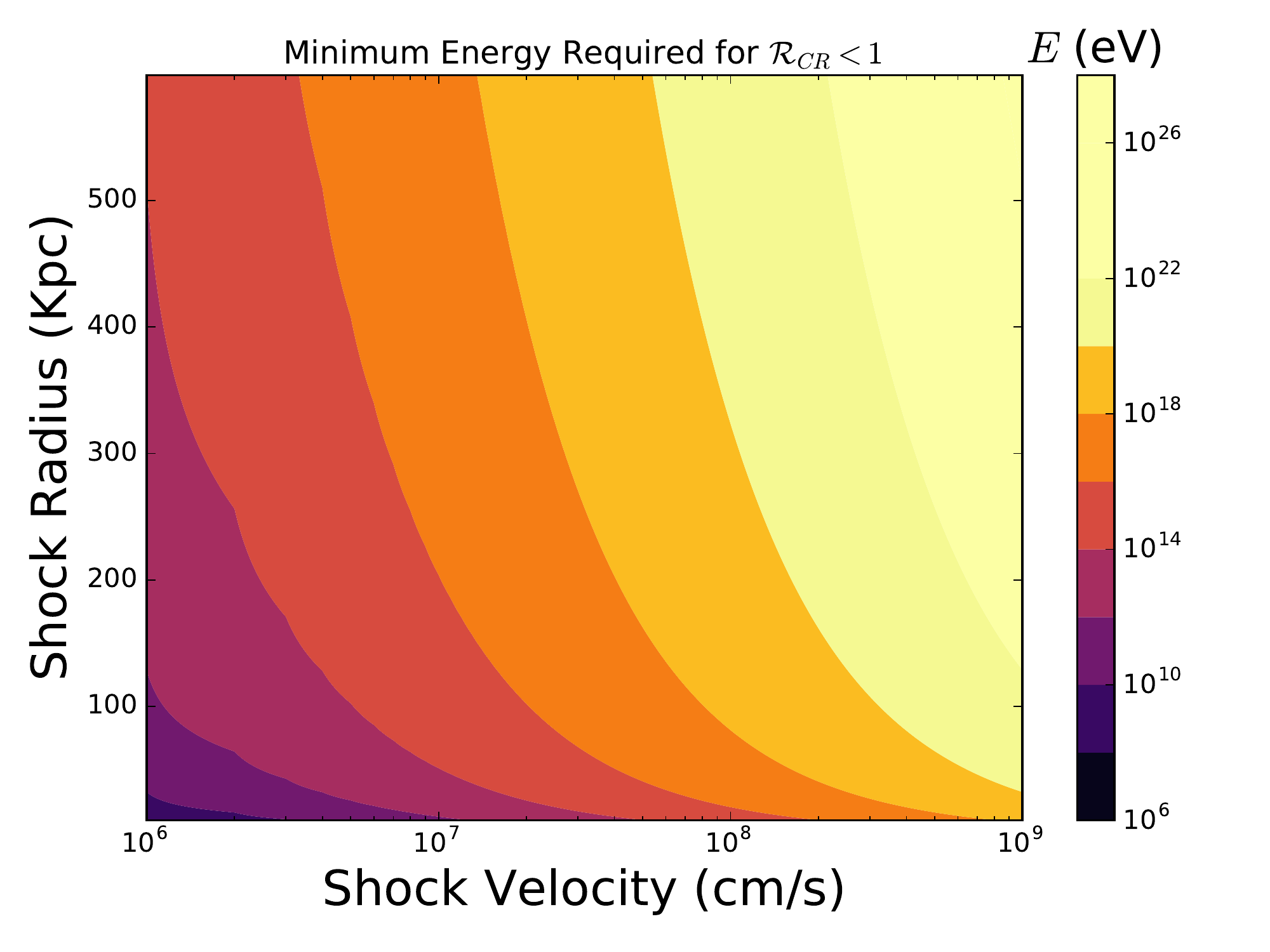}
\includegraphics[scale = 0.44]{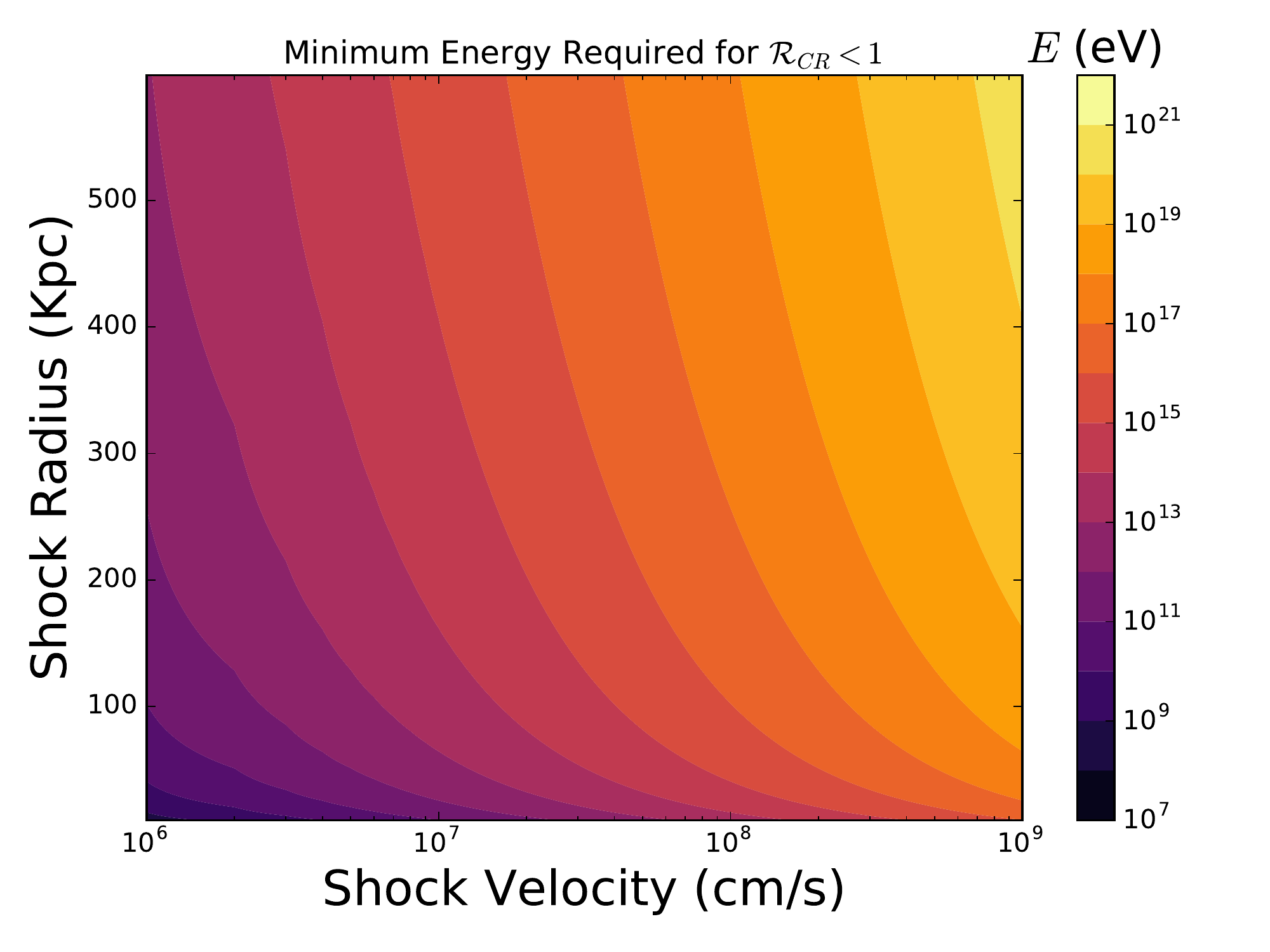}
\includegraphics[scale = 0.44]{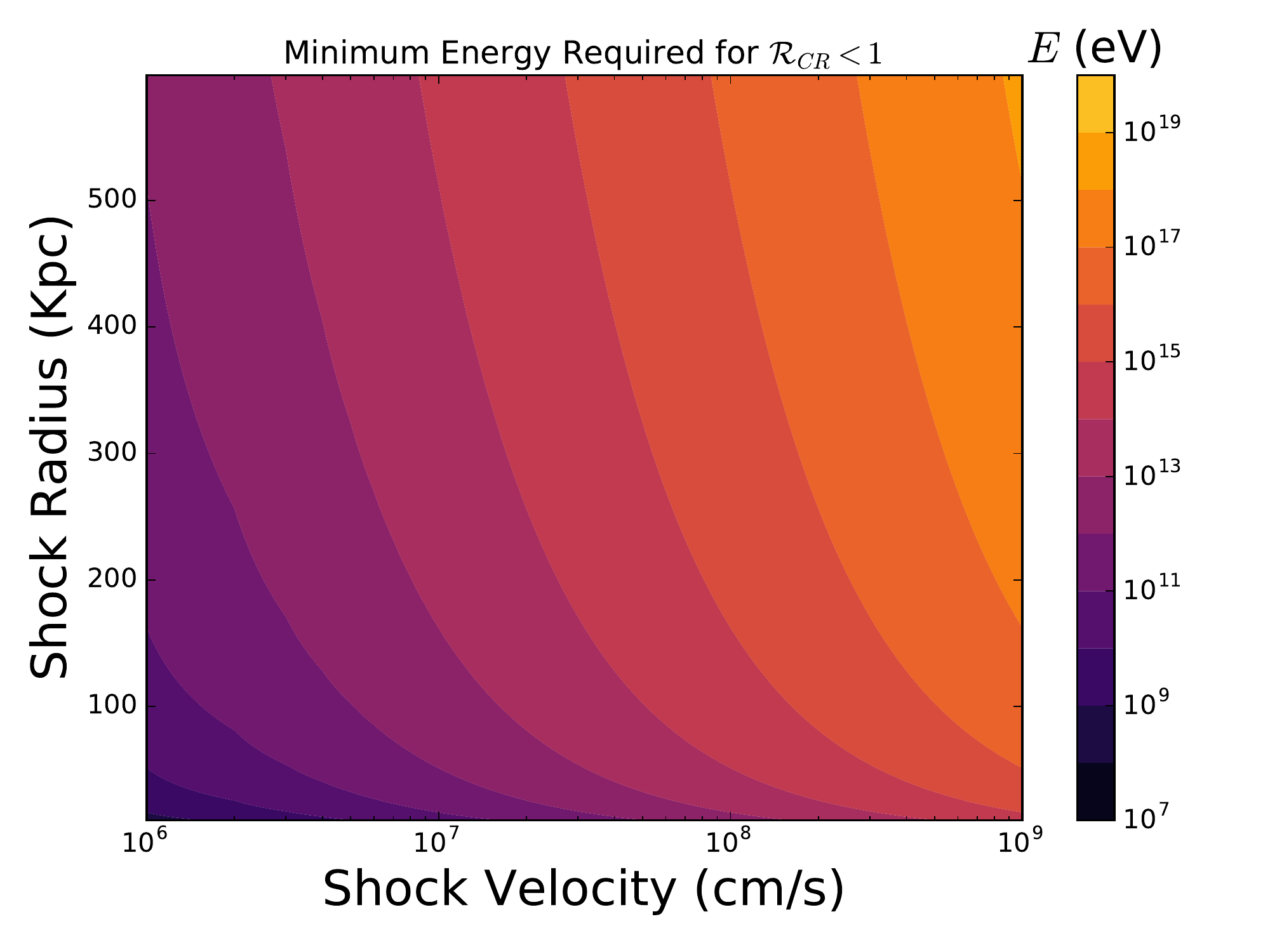}
\includegraphics[scale = 0.44]{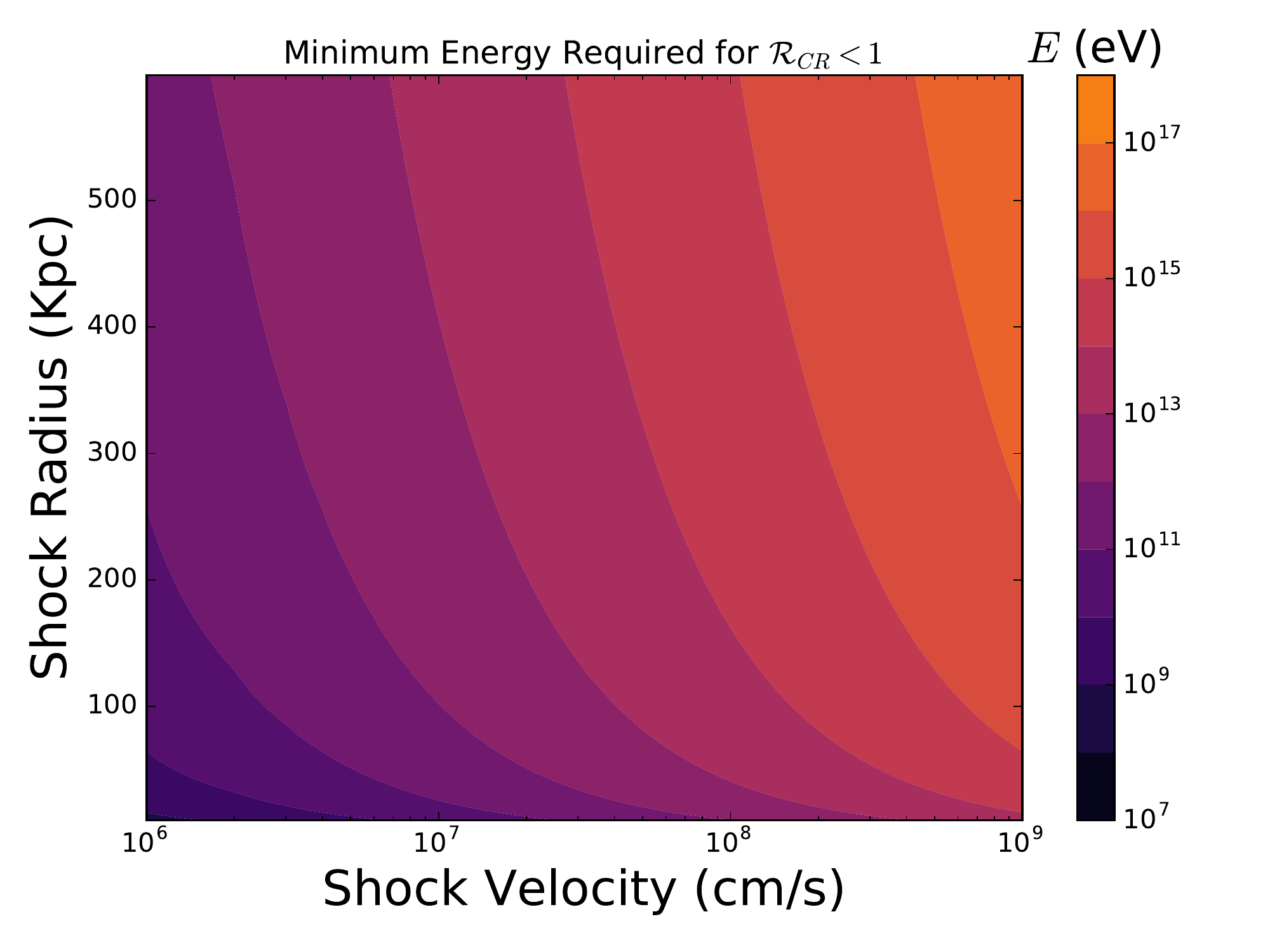}
\caption{Minimum energy required for accelerated CRs to be diffusion dominated, i.e. $\mathcal{R}_{CR} < 1$, for diffusion coefficients of $5 \times 10^{28} \rm cm^{2} s^{-1} E_{\rm GeV}^{a}$, with varying $a$. Top left: $a = 0.3$; Top right: $a = 0.4$; Bottom left: $a = 0.5$; Bottom right: $a = 0.6$. $t_{\rm acc} = 100$ Myrs. This plot is independent of choice for $P_{\rm IGM}$. For small $a$, the CR energy requirement for diffusion to dominate advection is quite high for a large range of velocity and radius. Generally, advection greatly dominates diffusion, specifically at the high velocities needed to accelerate CRs beyond the knee. For high $a$, however, the energy requirement is modest for most shock radii and velocities, boosting the chances of shin CRs diffusing back to the galaxy. Clearly, diffusion depends very sensitively on the diffusion coefficient used, making it difficult to draw conclusions on whether CRs will be able to diffuse back to the galaxy.}
\end{figure*}

\begin{figure}
\label{varyPIGM_reynoldsStrong}
\centering
\includegraphics[scale = 0.4]{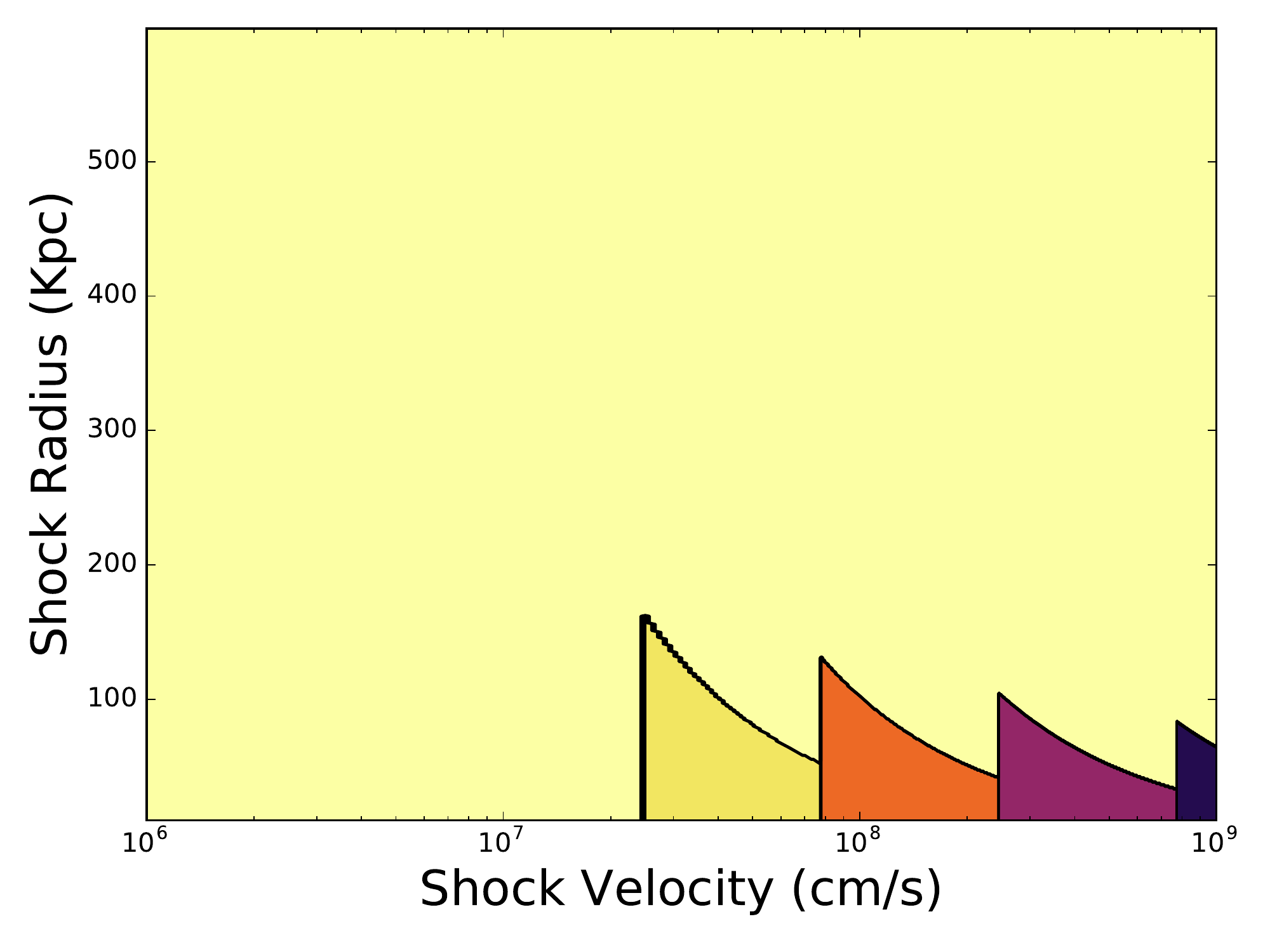}
\includegraphics[scale = 0.4]{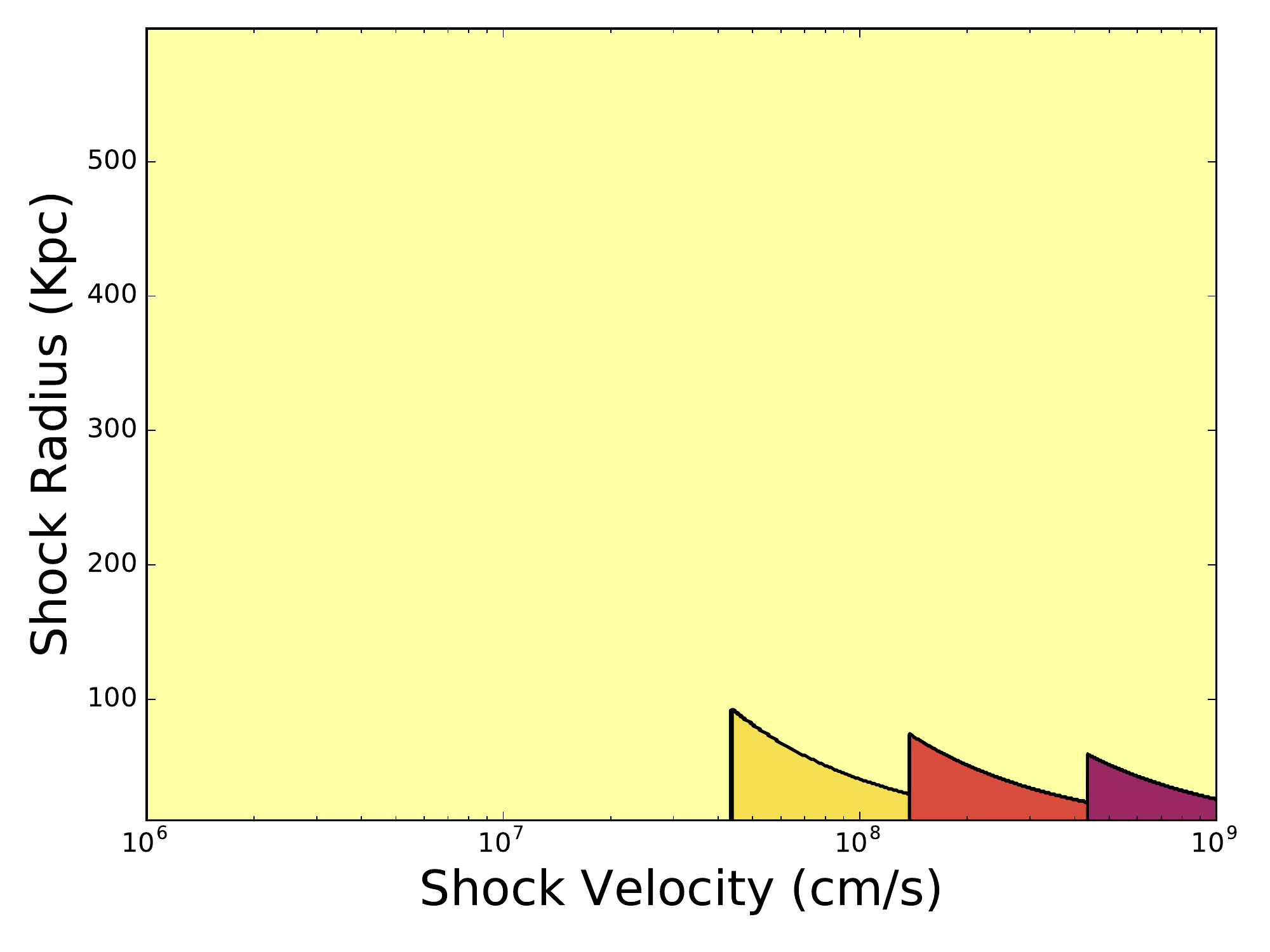}
\includegraphics[scale = 0.4]{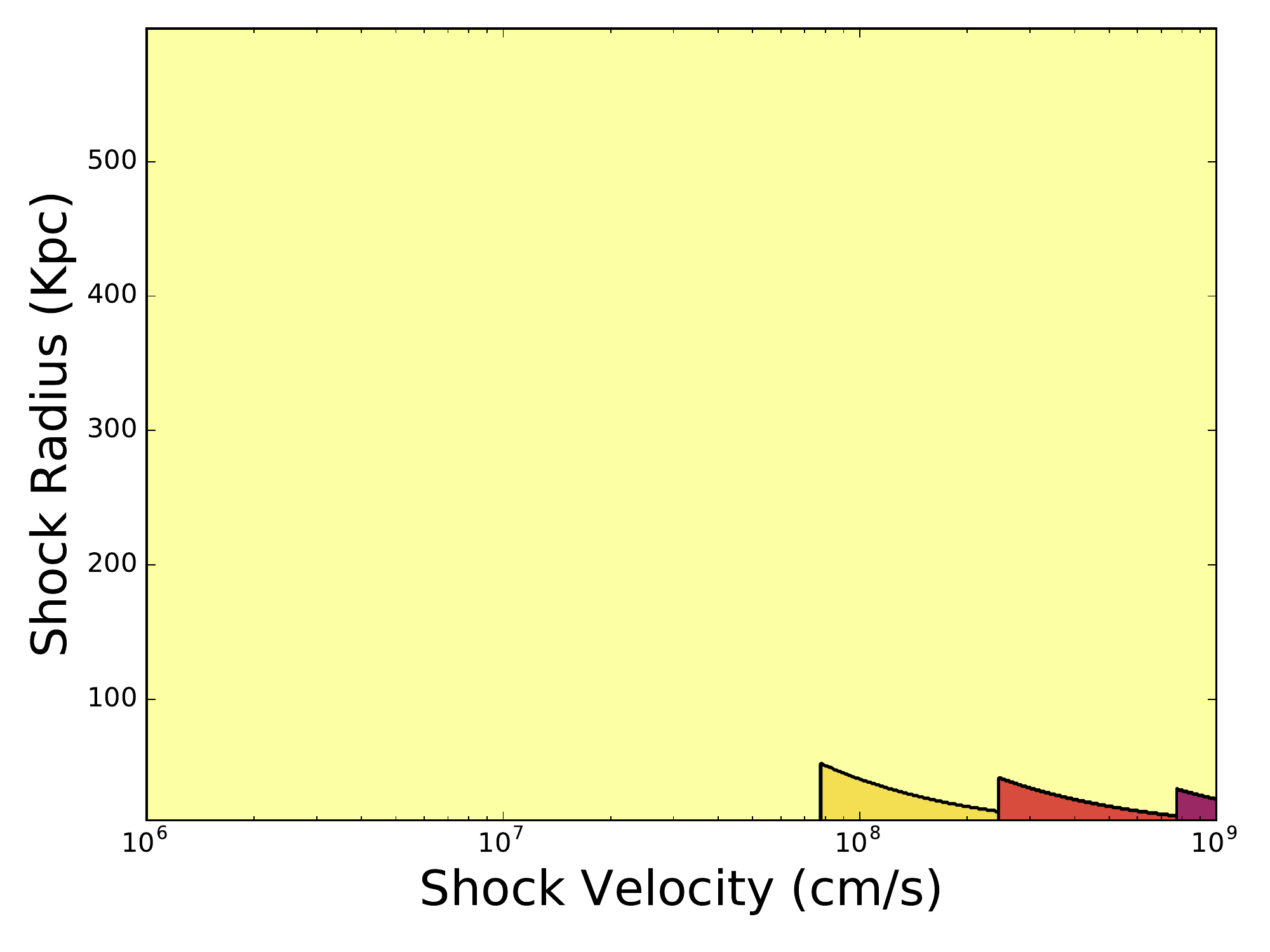}
\caption{Combining Figure \ref{reynoldsStrong_vs_Rshock_vs_Ushock_varya} with velocity requirements for CRs to be accelerated to energies of $10^{15}$ eV (yellow), $10^{16}$ eV (orange), $10^{17}$ eV (light purple), and $10^{18}$ eV (dark purple). High $P_{\rm IGM}$ are clearly favored to get low Reynolds numbers. At the same time, high $P_{\rm IGM}$ increase the CR acceleration rate, meaning lower velocities are needed to achieve the same CR energies. Top: $P_{\rm IGM} = 10^{-14}$ ergs $\rm cm^{-3}$; Middle: $P_{\rm IGM} = 10^{-15}$ ergs $\rm cm^{-3}$; Bottom: $P_{\rm IGM} = 10^{-16}$ ergs $\rm cm^{-3}$}
\end{figure}

$P_{\rm IGM}$ can also be eliminated in favor of the shock radius using Equation (\ref{shockradius}), and Equation (\ref{dEdtEquipartition}) can be rewritten in terms of the mass-loading $\dot{M_{\odot}}$ in units of $M_{\odot}/yr$, the shock velocity $V_{1000}$ in units of 1000 km/s, and the shock radius $R_{100}$ in units of 100 kpc. 

\begin{equation}
\label{usefuldEdt}
\frac{dE}{dt} = 0.12 Z \dot{M_{\odot}}^{1/2} V_{1000}^{5/2}/R_{100} \quad \text{GeV/yr}
\end{equation}

Let us now estimate whether these CRs can diffuse back to the galaxy by introducing a CR Reynolds number, $\mathcal{R}_{CR}$, which is the ratio of the diffusion time, $\tau_{\rm diff} = R_{\rm shock}^{2}/\kappa$, to the advection time, $\tau_{\rm adv} = R_{\rm shock}/V_{\rm shock}$, where $\kappa$ is the diffusion coefficient of CRs.  At the termination shock itself, where the CR flux and the corresponding wave excitation rates are very large, the Bohm diffusion limit is likely appropriate. This scenario occurs if the particle mean free path is equal to a particle gyroradius. In using the maximum acceleration rate of \cite{1983A&A...125..249L}, we have assumed Bohm diffusion. This limit, however, represents the minimum diffusion coefficient, meaning that CRs undergoing Bohm diffusion will have only a very small chance of diffusing back to the galaxy. 

For the bulk of the wind, however, diffusion coefficients estimated from galactic propagation models are likely more appropriate, and these coefficients can be orders of magnitude larger than those assuming Bohm diffusion. Typical values of the diffusion coefficient are given in \cite{2007ARNPS..57..285S}:

\begin{equation}
\label{StrongDiffusion}
\kappa(E) = D_{0} \times 10^{28} \rm cm^{2} s^{-1} E_{GeV}^{a}
\end{equation}
where $a$ varies from $0.3 - 0.6$ and $D_{0}$ varies from $3 - 5$. Recent work by \cite{2016arXiv160503111T} uses a more generous diffusion coefficient from \cite{2014A&A...567A..33T} with $D_{0}$ a factor of ten higher than in Equation (\ref{StrongDiffusion}). We restrict ourselves to the coefficients of \cite{2007ARNPS..57..285S}, and we choose $D_{0} = 5$ and vary $a$ from $0.3$ to $0.6$ to be general. Ultimately, our results are very sensitive to $a$.

The CR Reynolds number is then 
\begin{equation}
\label{reynolds}
\mathcal{R}_{\rm CR} = \frac{R_{\rm shock}V_{\rm shock}}{\kappa(E)}
\end{equation}

Assuming CRs diffuse according to Equation (\ref{StrongDiffusion}) and setting $\mathcal{R}_{\rm CR} = 1$, we can solve for the minimum CR energy, given a certain shock velocity and radius, such that CRs can diffuse back. Using $D_{0} = 5$, 

\begin{equation}
\label{Emin}
E_{\rm min, \rm GeV} = \Big( \frac{R_{\rm shock} V_{\rm shock}}{5 \times 10^{28}} \Big)^{1/a}
\end{equation}

From Equation (\ref{reynolds}), we can make a few statements: 1) CR diffusion is more favorable at lower shock radius. This is intuitive as CRs accelerated closer to the galaxy should have a greater chance of diffusing back to the galaxy. CR diffusion is consequently more favorable when $P_{\rm IGM}$ is higher because that pushes the shock radius inward. 2) Because $\mathcal{R}_{CR} \sim 1/E^{a}$, diffusion is more favorable for higher energy CRs. Figure \ref{reynoldsStrong_vs_Rshock_vs_Ushock_varya} considers winds of various shock radius and velocity and plots the minimum CR energy required for CRs to diffuse back to the galaxy as given in Equation (\ref{Emin}). Figure \ref{reynoldsStrong_vs_Rshock_vs_Ushock_varya} clearly shows that CRs of lower energies can diffuse back to the galaxy more easily if the shock has a small velocity and small radius. In general, higher energy CRs diffuse back more easily; however, whether shocks of a certain velocity can accelerate CRs to high energies depends on the shock velocity, as seen in Figure \ref{maxdEdt}. Figure \ref{varyPIGM_reynoldsStrong} combines the upper right panel ($a = 0.4$) of Figure \ref{reynoldsStrong_vs_Rshock_vs_Ushock_varya} with the velocity requirements shown in Figure \ref{maxdEdt} to show the regions in $R_{\rm shock} - V_{\rm shock}$ plane in which CRs may be accelerated to a certain energy and also diffuse back to the galaxy. Note that, for each CR energy, those CRs can only satisfy these requirements within a certain band of shock velocities. If velocities are too low, the shock will not be able to accelerate CRs to that energy. If the velocity is too high, the CR Reynolds number will be too high due to strong advection with the flow. 

One can also see from Figure \ref{reynoldsStrong_vs_Rshock_vs_Ushock_varya} how the picture we have constructed will change greatly depending on the choice of energy exponent, $a$. For $a = 0.3$, almost no wind in the $R_{\rm shock} - V_{\rm shock}$ plane will have $\mathcal{R}_{CR} < 1$, whereas for $a = 0.6$, a majority of shin CRs will be diffusion dominated.

\section{Particle Acceleration: Necessary Conditions in the Galaxy}
\label{galaxyConditions}

\begin{figure*}
\label{T0_vs_ShockVel_vs_Beta}
\centering
\includegraphics[scale = 0.445]{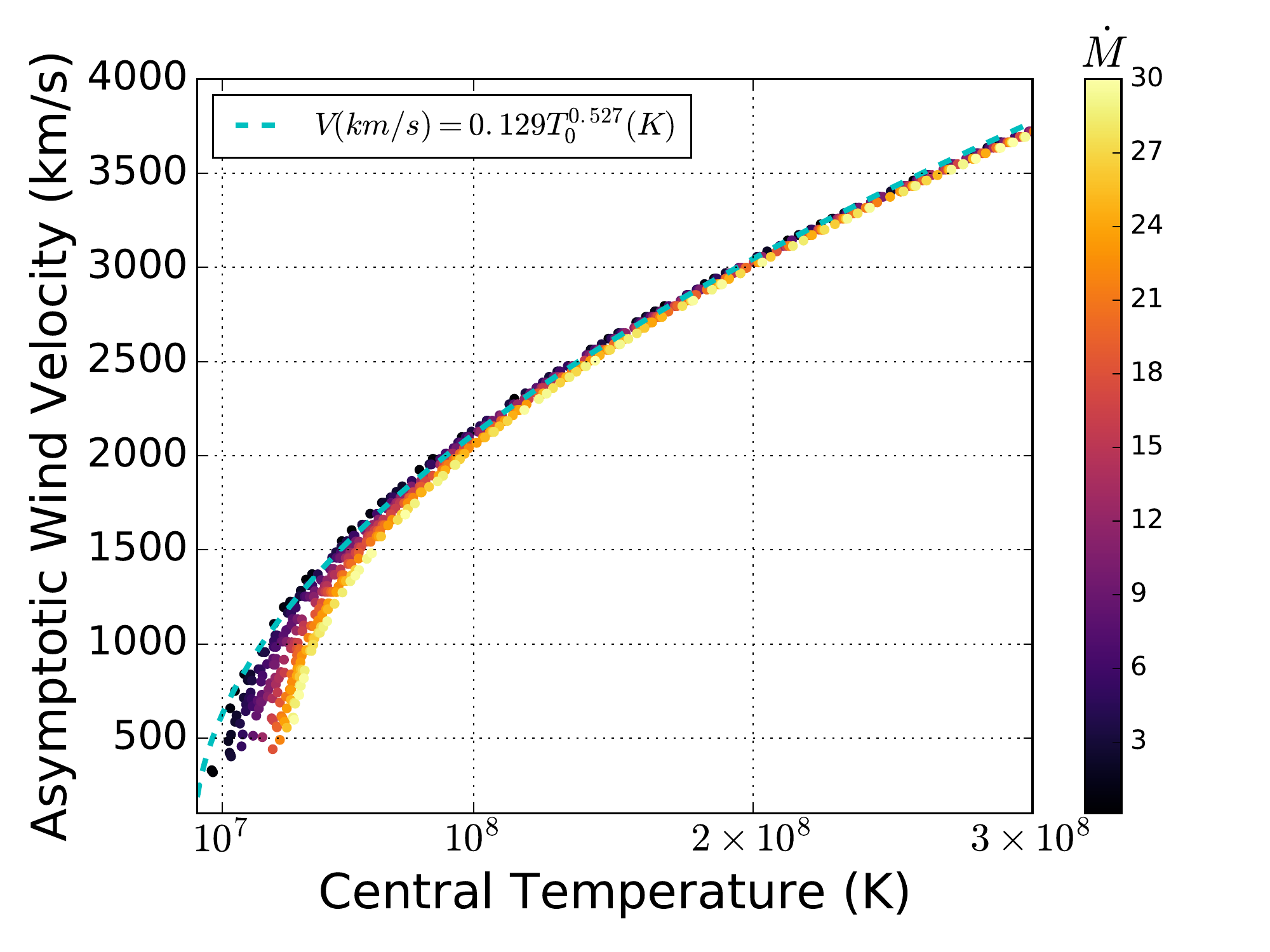}
\includegraphics[scale = 0.425]{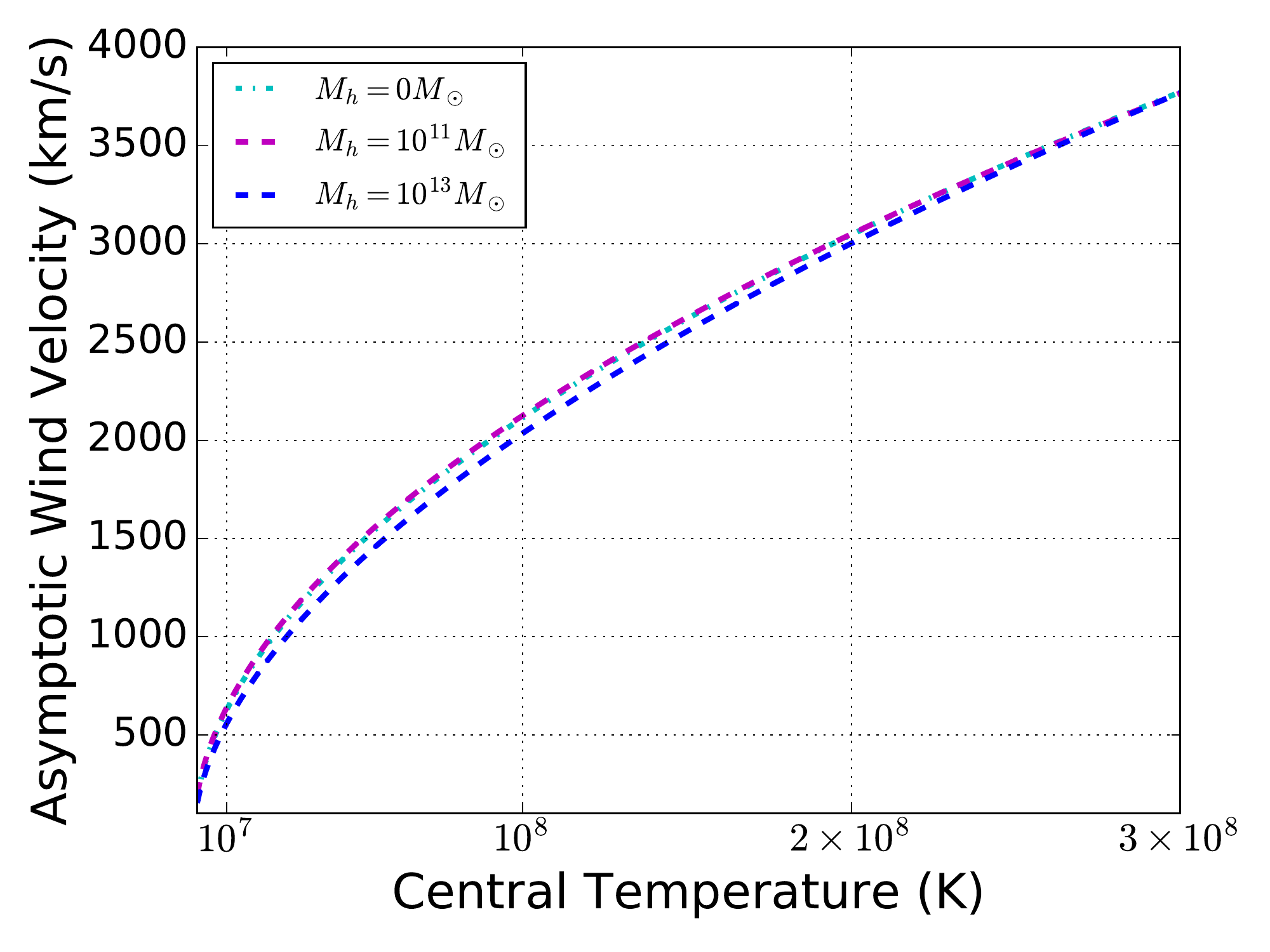}
\caption{Left: Plot of asymptotic wind velocity vs. central temperature vs. mass-loading in solar masses per year for a set of wind solutions from a $10^{9} M_{\odot}$ galaxy. At low temperatures (low initial thermal energy budget), highly mass-loaded outflows are affected by radiative losses, resulting in higher initial temperatures needed to drive an outflow of a certain velocity. At high temperatures (high initial energy budget), winds are blown out at high velocities with gravity and radiative losses both playing a negligible role. At these high velocities, there is very little dependence of velocity on $\dot{M}$. The function $V_{\rm shock} (\rm km/s) = 0.129T_{0}^{0.527} (K)$ is a best fit to the $\dot{M} < 2$ winds and is plotted to show the dependence at high velocities. Right: Best fit lines for same wind model runs but with halo masses of $10^{11}M_{\odot}$ and $10^{13}M_{\odot}$ and corresponding values of $6$ kpc and $25$ kpc, respectively, for the scale length of the halos. Halo mass seems to have only a small effect on the wind dynamics, especially for winds with high velocities.}
\end{figure*}

\begin{figure*}
\label{Rshock_contour}
\centering
\includegraphics[scale = 0.44]{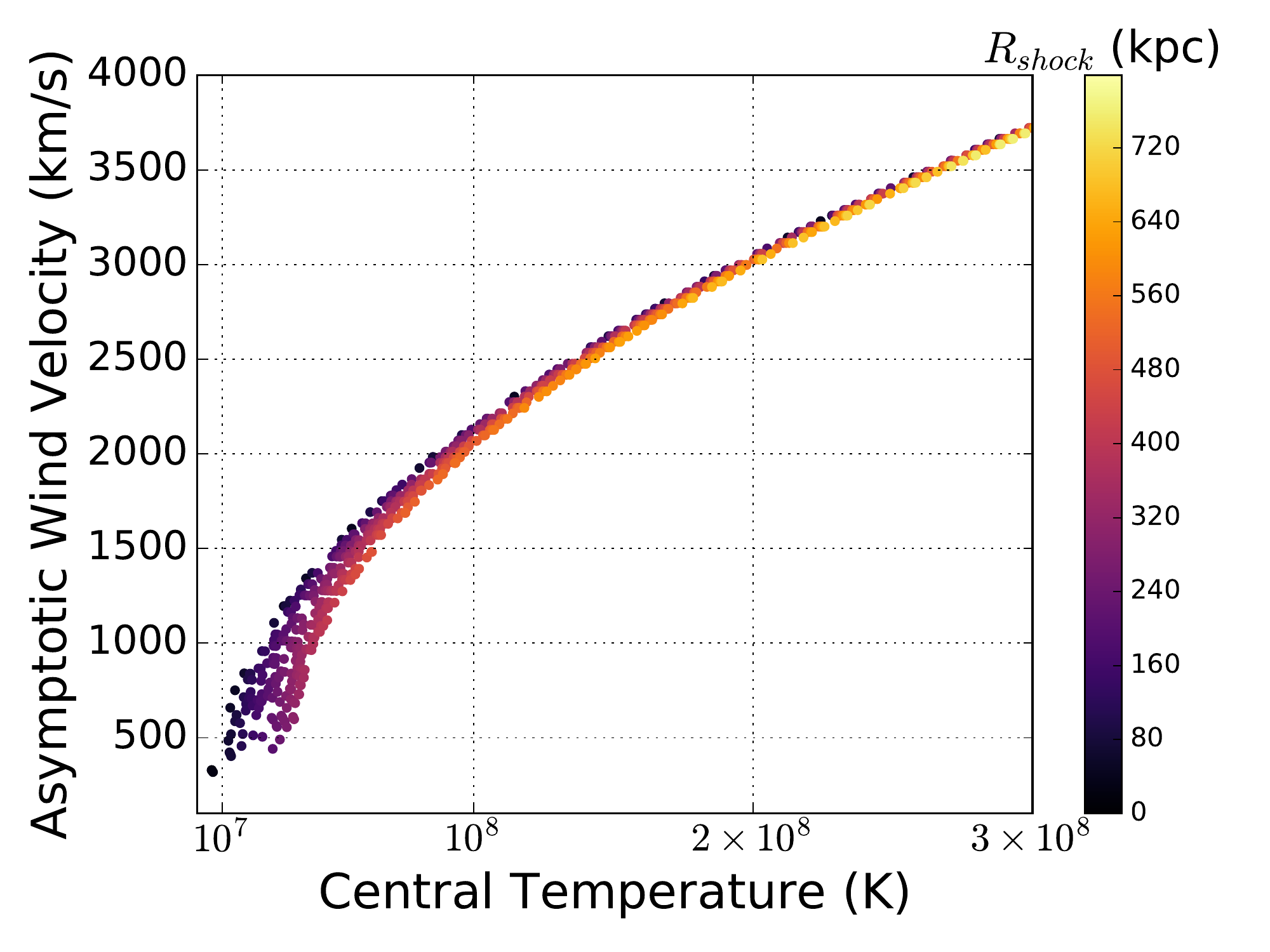}
\includegraphics[scale = 0.44]{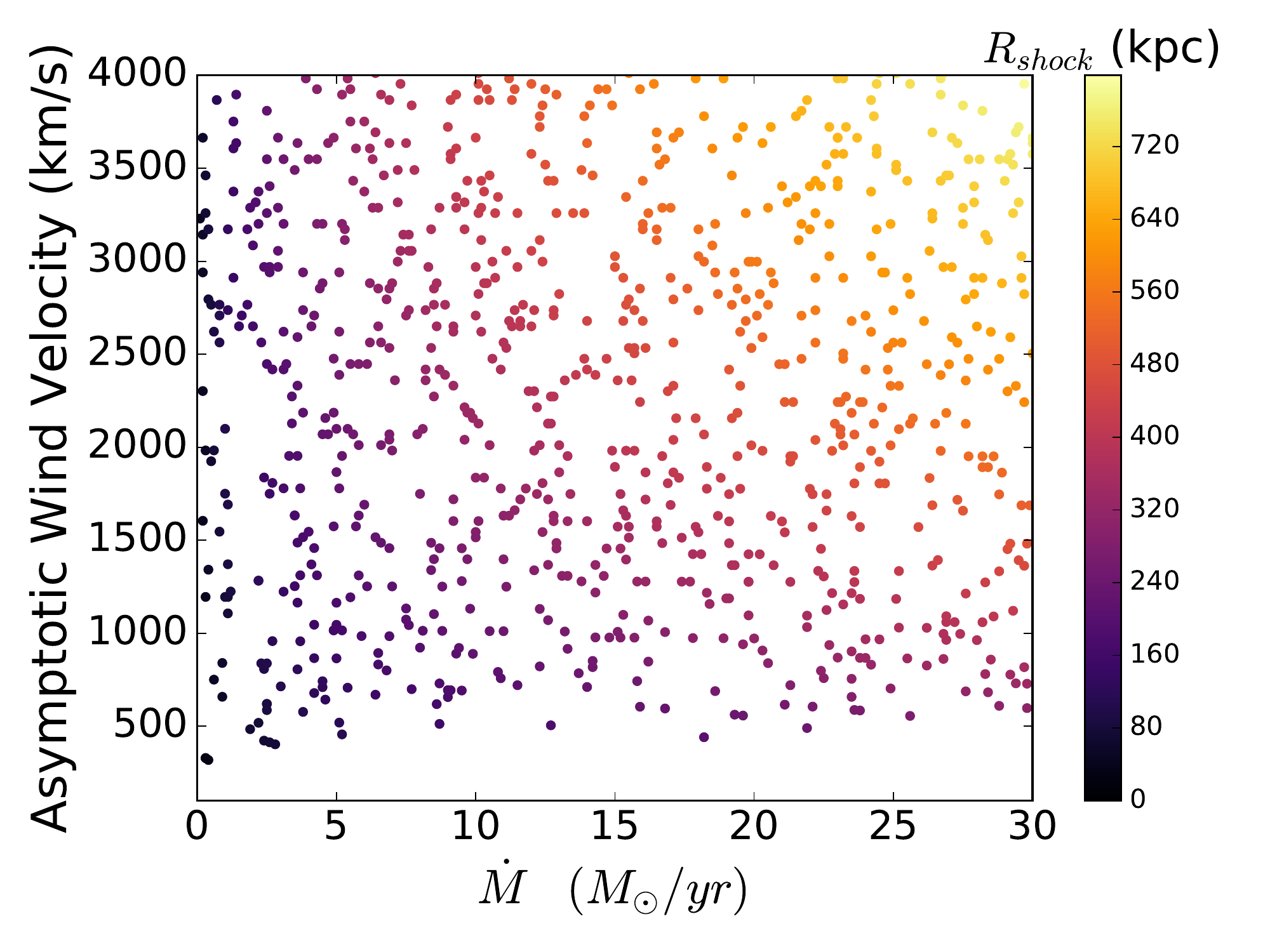}
\caption{Varying $\dot{M}$ and $T_{0}$ (effectively varying $V_{\rm shock}$), we plot the shock position for a number of wind models run. Shocks occur close to the galaxy when the wind velocity and $\dot{M}$ are both low, as can be explicitly seen in Equation (\ref{shockradius})}
\end{figure*}

Up to this point, no connection has been made to our specific model described in Section \ref{windModel}. We have only made assumptions on the magnetic field (amplified to equipartition) and the form of the CR diffusion coefficient (given in Equation (\ref{reynolds})). Let us now run a set of wind simulations and see where the winds shock and at what velocity. We choose to analyze a $10^{9} M_{\odot}$ galaxy, and we vary $\dot{M}$ and the central temperature (i.e. the thermal energy budget) of the outflow without any additional energy addition from supernovae (i.e. $\alpha = 0$). We set $P_{\rm IGM} = 10^{-14} \rm ergs$ $ \rm cm^{-3}$.

First, it is informative to see what type of energy budget is required to achieve winds of various velocities.  The resulting plot, Figure \ref{T0_vs_ShockVel_vs_Beta}, shows a clear outflow velocity dependence on $\beta$ for low central wind temperatures, $T_{0}$. For higher mass-loaded winds that are more susceptible to radiative energy losses, higher central temperatures are required to achieve the same asymptotic wind velocity as lower mass-loaded outflows. At higher central temperatures, the wind temperature is further from the peak of the cooling curve, and the outflow greatly overcomes gravity. These outflows are now akin to the high velocity outflows of the gravity-less \cite{1985Natur.317...44C} model, and we see that the asymptotic velocity now has virtually no dependence on mass-loading factor. In fact, the dependence roughly follows the curve $V_{\rm shock} (\rm km/s) = 0.129T_{0}^{0.527} (K)$. For comparison, the central sound speed is $c_{s0} (\rm km/s) = \sqrt{\gamma k_{B} T/m} \approx 0.15 T_{0}^{0.5} (K)$. Therefore, the asymptotic velocity closely, but not exactly, follows the central sound speed, neglecting radiative effects. 

\begin{equation}
\label{velCS}
V_{\rm shock} \approx c_{s0}
\end{equation}
This is physically intuitive because the central thermal energy is converted almost entirely to kinetic energy. Table \ref{windConditionsTable} in the Appendix gives the necessary wind velocities, for various acceleration times, required to accelerate $10^{15}$, $10^{16}$, $10^{17}$, and $10^{18}$ eV CRs, as well as the central temperatures required to achieve these wind velocities. The minimum wind velocity is independent of wind model, whereas the central temperature is specific to our wind model.

We also generated wind solutions using different halo masses. Whereas the left panel of Figure \ref{T0_vs_ShockVel_vs_Beta} assumes a halo mass $M_{h} = 0$, the right panel shows the best fit lines for various halo masses. We can see immediately that the added halo mass does not affect the model significantly. The largest difference is that the trend for the $10^{13} M_{\odot}$ halo is shifted downward when compared to the model without the halo. This implies that massive halo winds need slightly higher central temperatures for the same asymptotic wind velocity and mass-loading as non-halo winds. One explanation for the minimal effect of the added halo is that there is very little halo mass inside the critical radius.

\begin{figure*}[!h]
\label{T0_vs_ShockVel_vs_Emax}
\centering
\includegraphics[scale = 0.44]{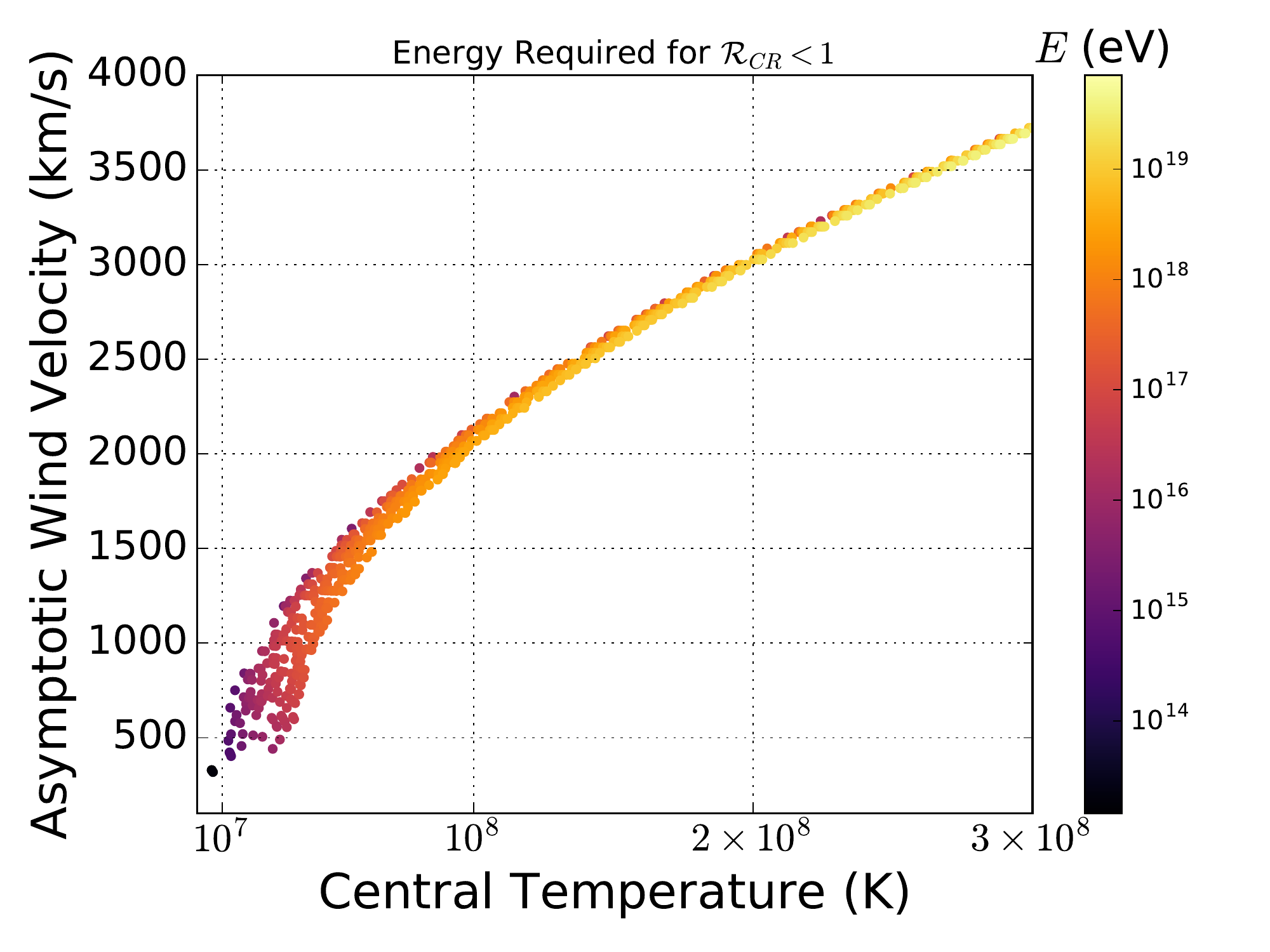}
\includegraphics[scale = 0.44]{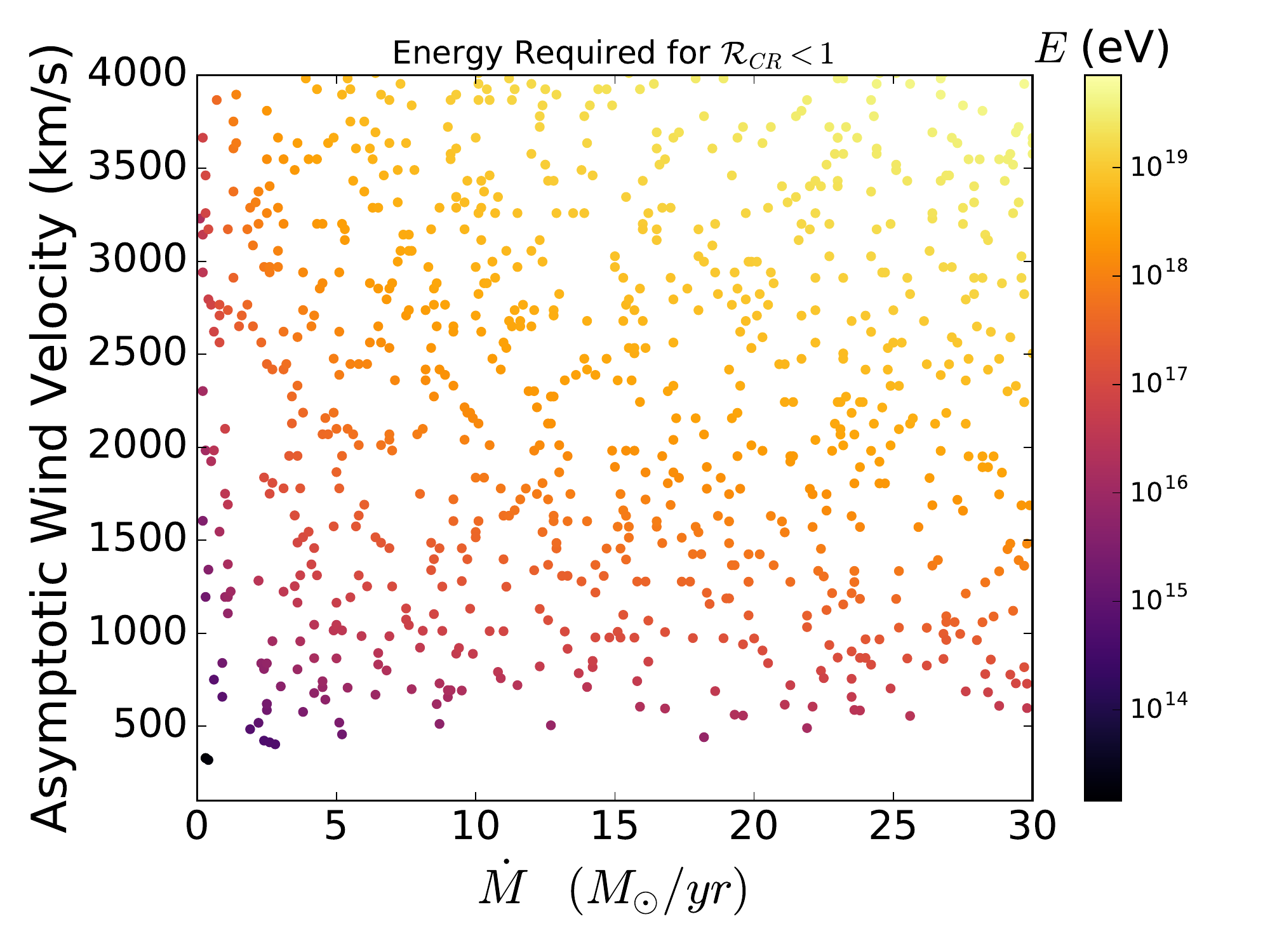}
\includegraphics[scale = 0.44]{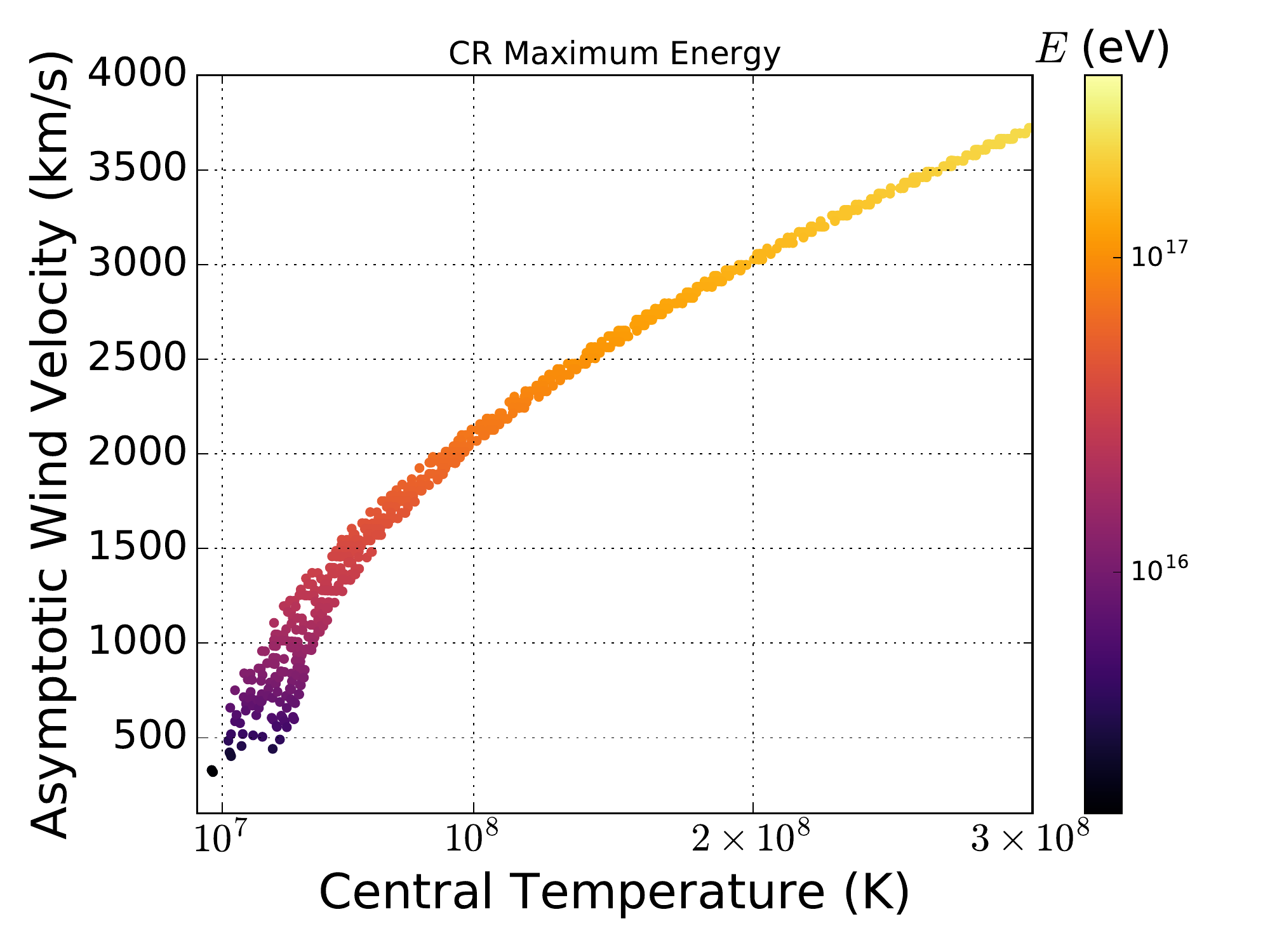}
\includegraphics[scale = 0.44]{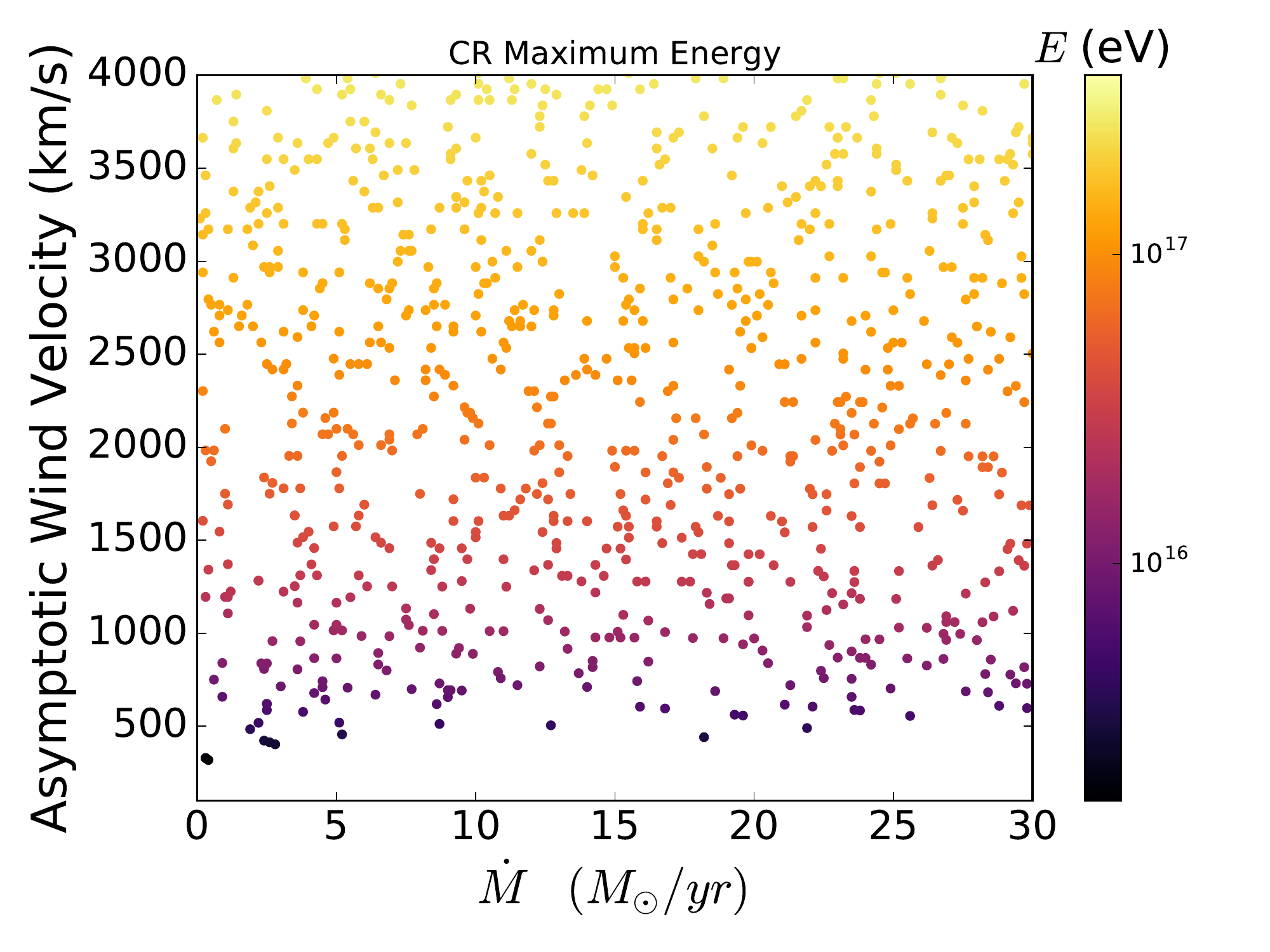}
\includegraphics[scale = 0.44]{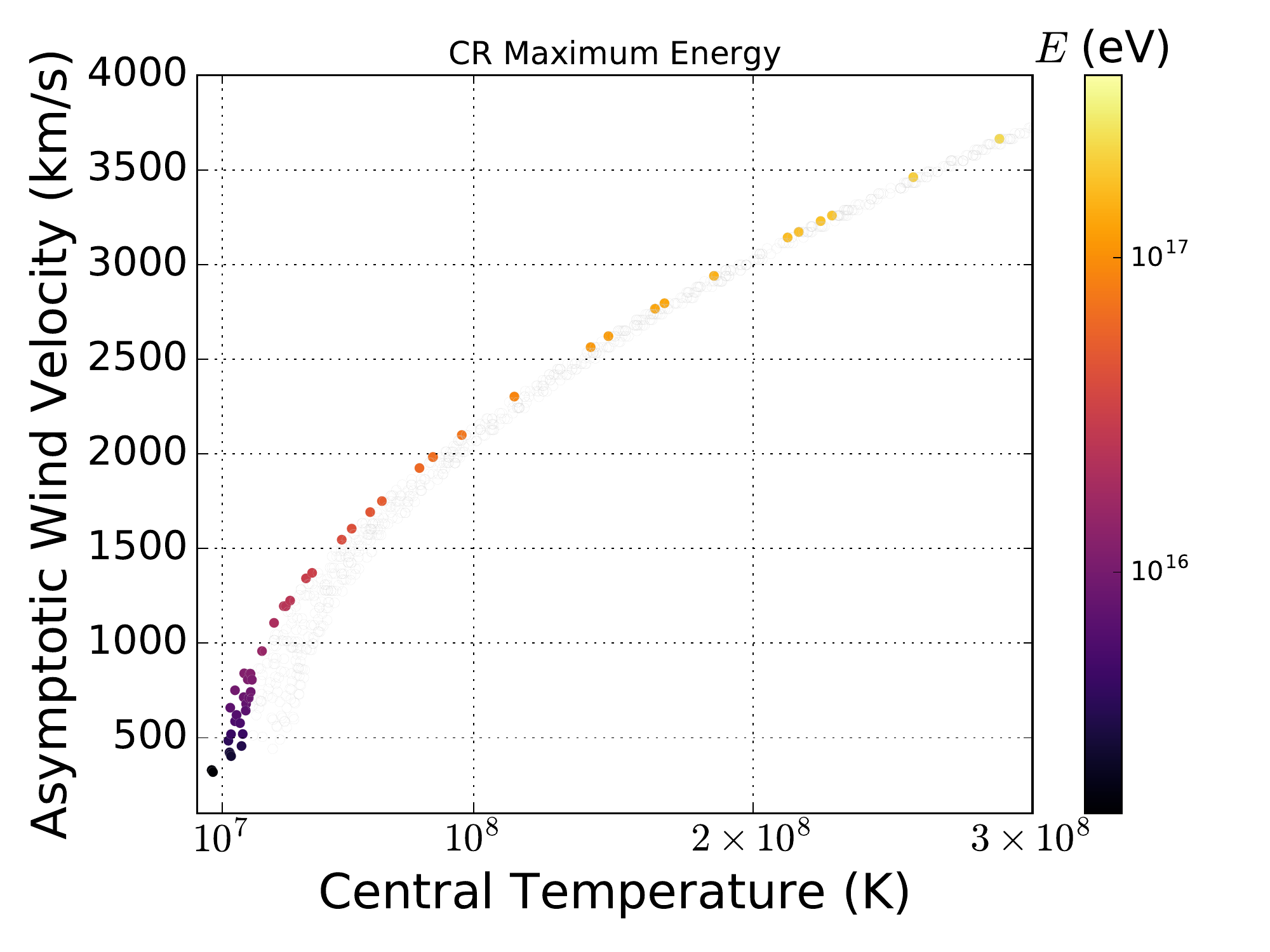}
\includegraphics[scale = 0.44]{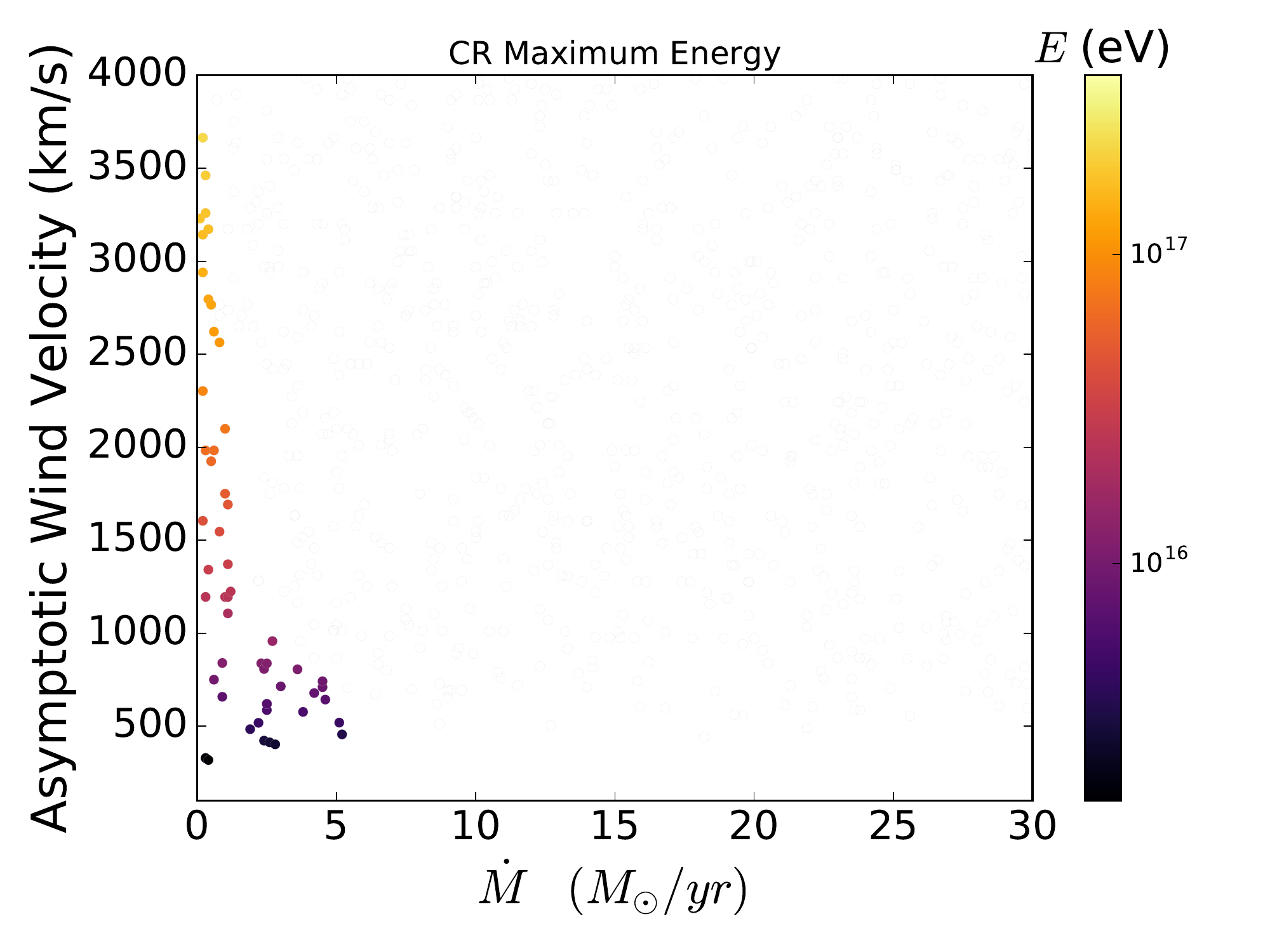}
\caption{Top: The minimum energy required for CRs to diffuse back to the galaxy, assuming $a = 0.4$ in Equation (\ref{StrongDiffusion}). Middle: Maximum energy of CRs accelerated by wind termination shocks after $t_{\rm acc} = 100$ Myrs. Bottom: The maximum energy of accelerated CRs, but only those CRs that are also able to diffuse back. If the maximum energy that can be achieved is less than the minimum energy required to have $\mathcal{R}_{CR} < 1$, then no CRs of any energy generated by those shocks will be able to diffuse back to the galaxy. These wind solutions are shown as gray circles. The right panels show more explicitly the dependence on mass-loading. For winds with lower asymptotic velocities and lower mass-loading, the minimum energy is lower, resulting in a higher fraction of CRs that can diffuse back to the galaxy. }
\end{figure*}

Now that we have addressed what input parameters are needed to achieve certain velocities, we will address the issue of CR diffusion back to the galaxy.  The resulting shock positions for our sample of outflows are given in Figure \ref{Rshock_contour}. As expected, the shock position, which is effectively set by $\rho V_{\rm shock}^{2} = P_{\rm IGM}$, is further from the galaxy when more mass is loaded into the wind (greater $\dot{M}$) and when the shock velocity is higher. Implications for diffusion are shown in Figure \ref{T0_vs_ShockVel_vs_Emax}, which shows the maximum achievable CR energy compared to the minimum energy required for CRs to diffuse back to the galaxy given $a = 0.4$. In the bottom panel, winds for which no CRs can diffuse back, i.e. for which the maximum energy that can be achieved is less than the minimum energy required to have $\mathcal{R}_{CR} < 1$, are plotted as gray circles.  Among the many wind solutions shown, only a handful can produce CRs of any energy that can diffuse back to the galaxy. Holding energy and velocity constant, lower mass-loaded outflows shock closer to the galaxy, decreasing $\mathcal{R}_{CR}$. In the wind sample shown, only outflows with $\dot{M} \lesssim 5$ yield $\mathcal{R}_{CR} < 1$. Although CRs of energies between $10^{17}$ and $10^{18}$ eV will more easily diffuse back for a given shock velocity, they can only be produced in very high velocity shocks, which increases their chances of being advected with the flow. The results shown here seem to favor CR acceleration and subsequent diffusion only for energies between the knee and $10^{17}$ eV and for winds with preferentially low mass-loading factors and velocities.

\section{Cosmic Ray Luminosity}
\label{lums}

\begin{figure*}
\label{CRProduced_contour}
\centering
\includegraphics[scale = 0.44]{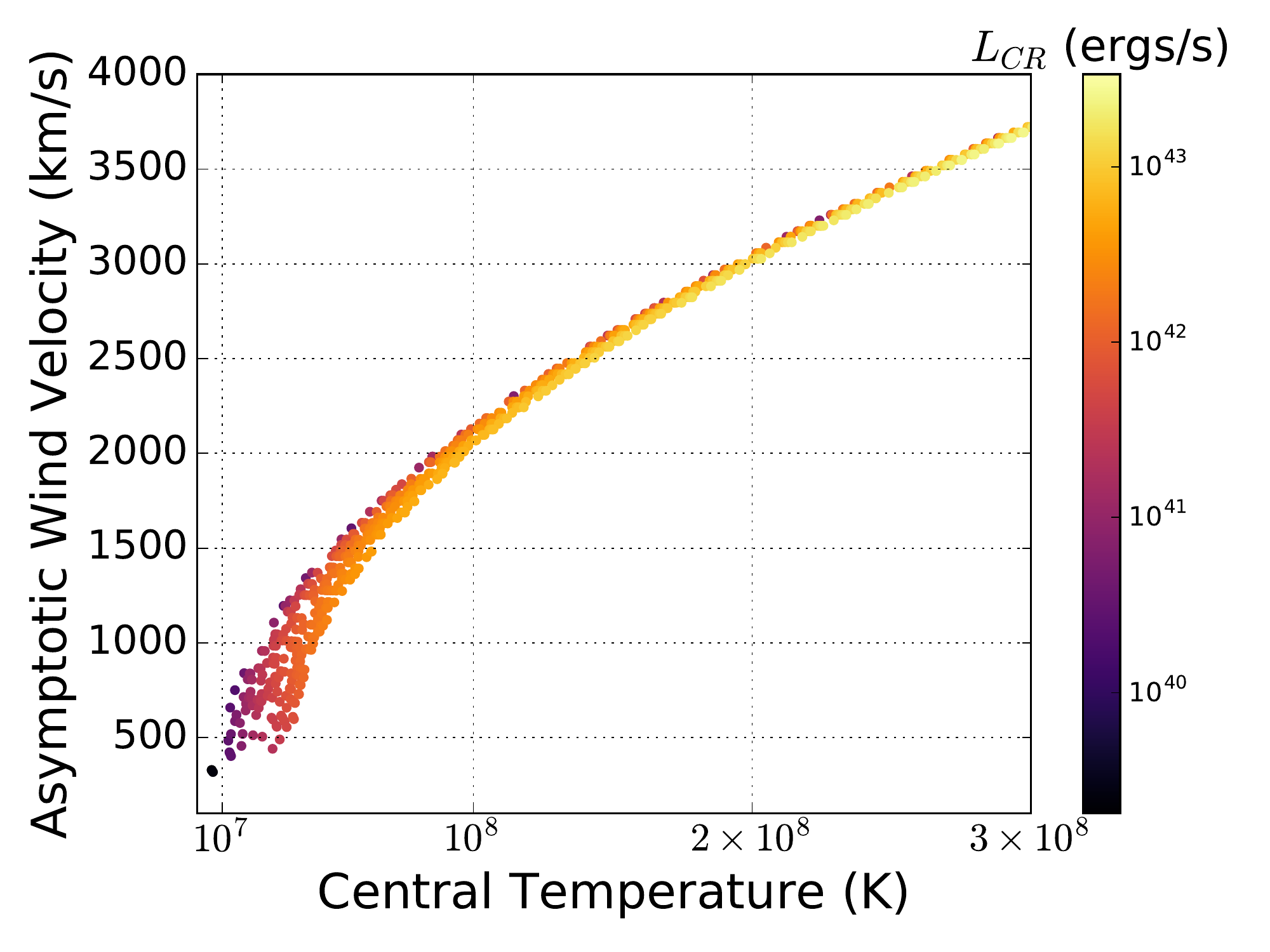}
\includegraphics[scale = 0.44]{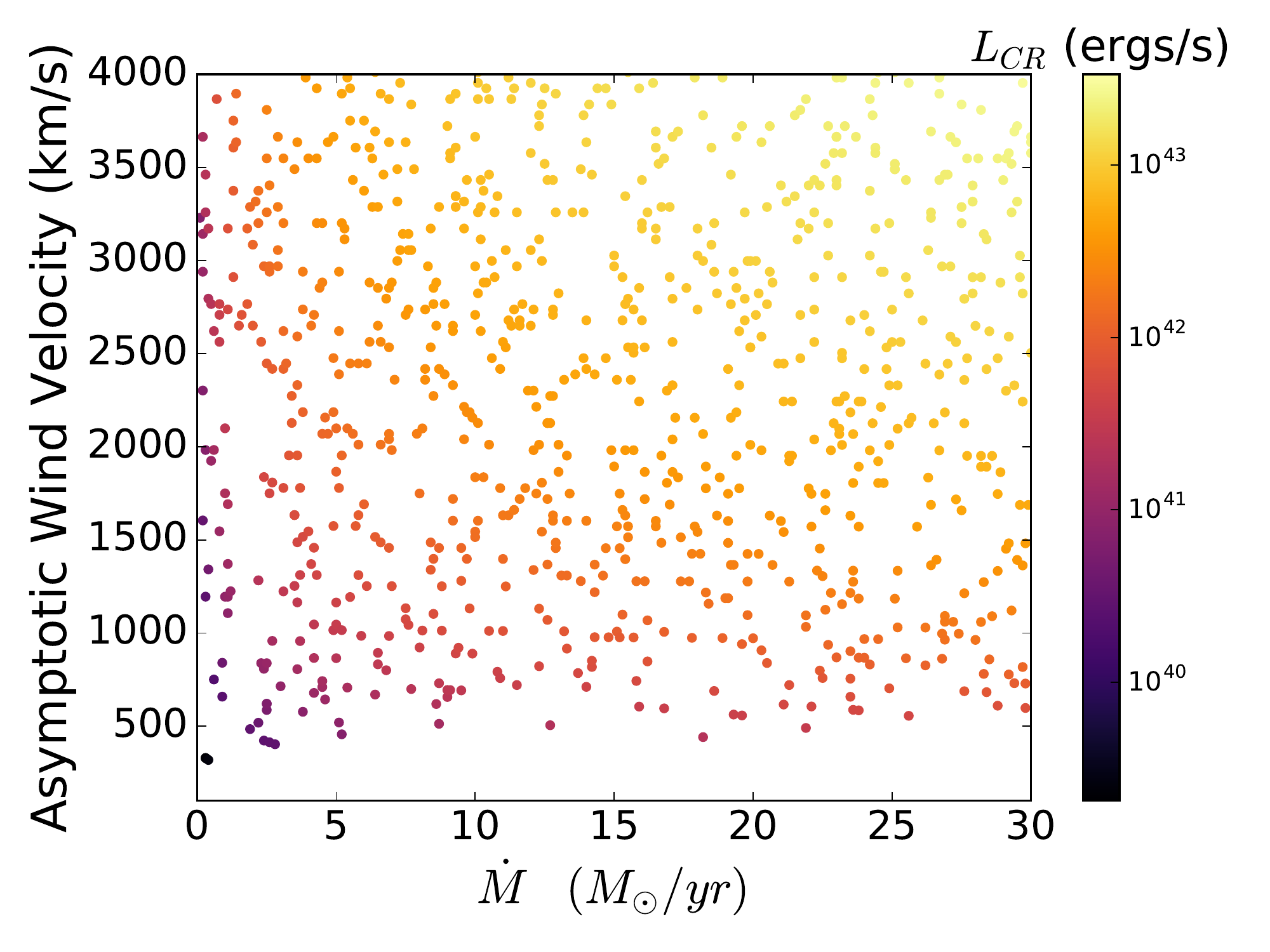}
\includegraphics[scale = 0.44]{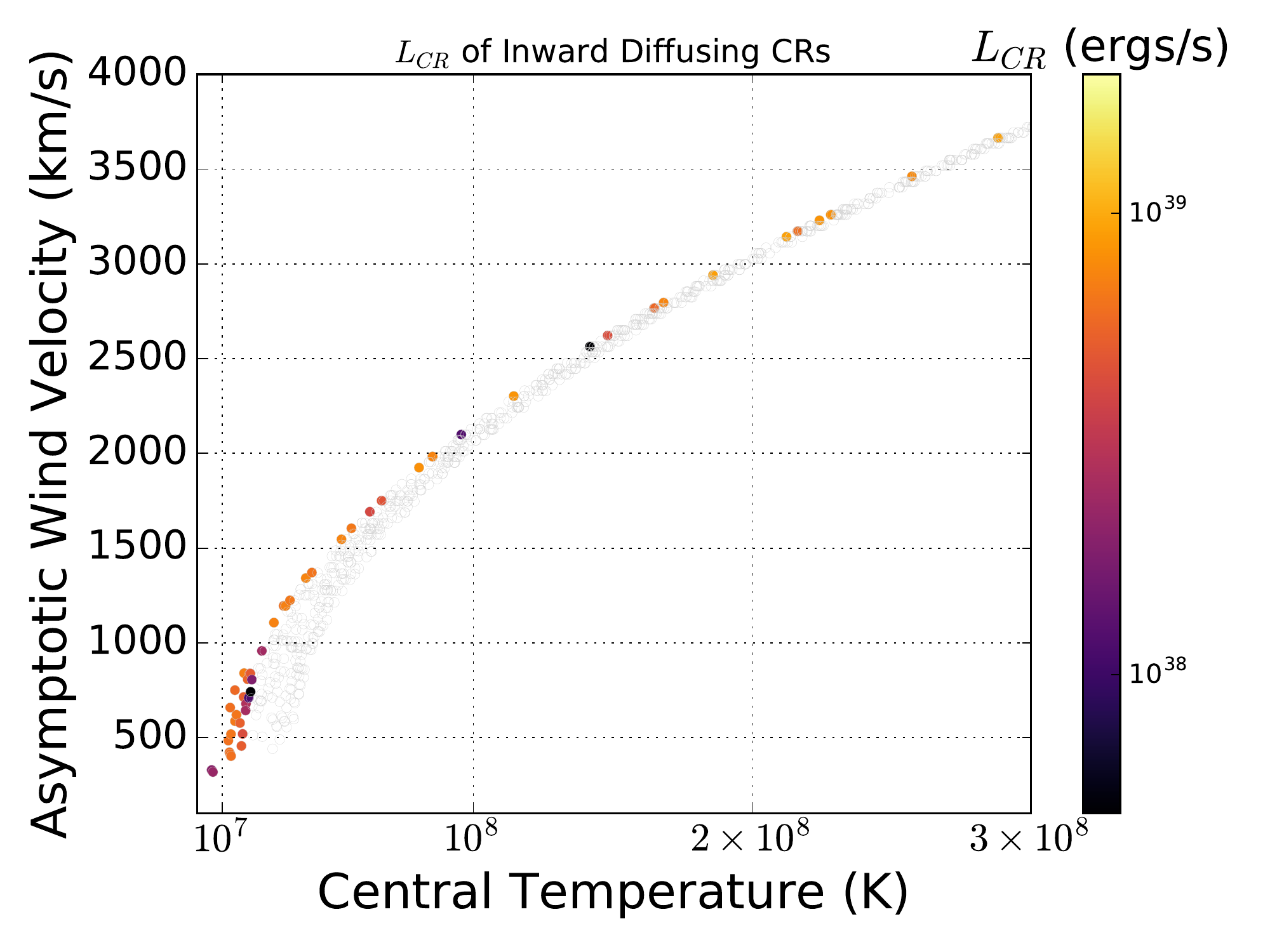}
\includegraphics[scale = 0.44]{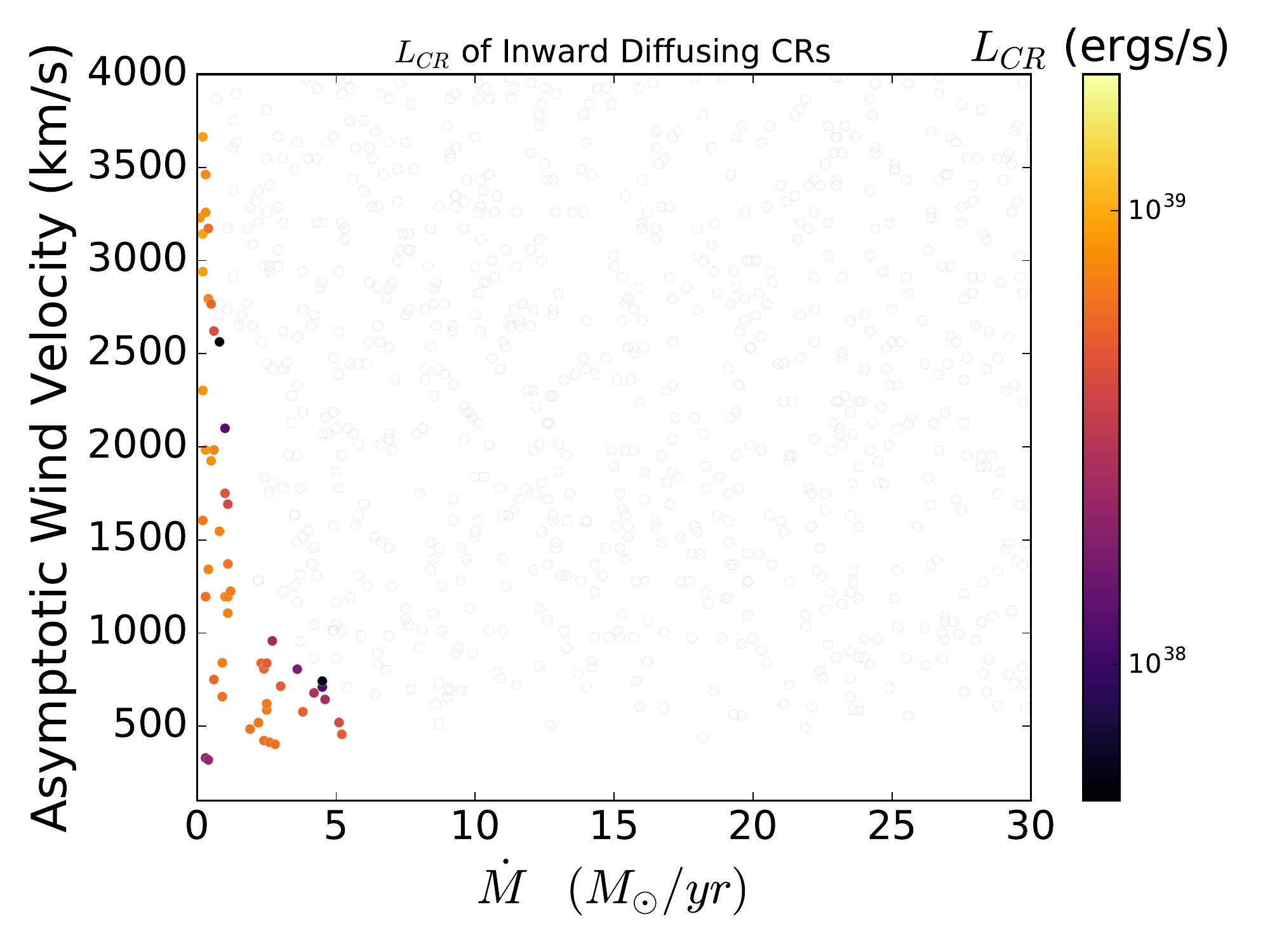}
\caption{Top: Total CR luminosity for varying $\dot{M}$ and $T_{0}$ (effectively varying $V_{\rm shock}$). For shocks with velocities greater than 1000 km/s, the CR luminosity varies from $10^{41} - 10^{43}$ ergs/s. Bottom: Luminosity of just the CRs that may diffuse back to the galaxy, assuming $a = 0.4$. Typically, the CRs that diffuse back account for a fraction of $10^{-3} - 10^{-1}$ of the total luminosity of shock accelerated CRs, resulting in luminosities that range from $10^{37} - 10^{39}$ ergs/s.}
\end{figure*}

In this section, we estimate the total CR production from a spherical wind termination shock. We note that $\rho V_{\rm shock}^{3}$ gives the energy flux of the shock, and we assume that some fraction, $\eta$, of that energy will be given to CRs. Integrating over a spherical shell at radius $R_{\rm shock}$, the total luminosity of CRs produced at the termination shock is  
\begin{equation}
\label{CRProduction}
L_{CR} = 4 \pi \eta R_{\rm shock}^{2} \rho V_{\rm shock}^{3} \approx \eta \dot{M}_{\odot} V_{\rm shock}^{2}
\end{equation}

The total CR luminosity for our sample of wind runs is plotted in the top panel of Figure \ref{CRProduced_contour}, assuming a tenth of the shock energy is converted to CR energy, i.e. $\eta = 0.1$. For shock velocities greater than $1000$ km/s, the total CR luminosity can be as high as $10^{43}$ ergs/s for highly mass-loaded outflows. This is $\approx$ 10 \% of the Milky Way luminosity, for comparison, suggesting that the re-acceleration of adiabatically expanded CRs at galactic wind termination shocks could be a significant source of energy flux into the IGM. This luminosity, however, is the \emph{total} luminosity in CRs, most of which are advected away with the flow. To estimate the luminosity of only the CRs that diffuse back to the galaxy, we assume the CR spectrum follows a power law from $E_{0} = 1$ GeV to $E_{\rm max}$ with the normalization of the total CR flux given by $K_{p}$. 

\begin{equation}
\eta \rho V_{\rm shock}^{3} = K_{p} \int_{E_{0}}^{E_{\rm max}}\Big(\frac{E}{E_{0}} \Big)^{-p} E dE
\end{equation}
Carrying out the integral and assuming $E_{\rm max}$ is much greater than $E_{0}$ (effectively taking $E_{\rm max}$ to be infinite), we find
\begin{equation}
K_{\rm p} = \frac{\eta \rho_{\rm shock} V_{\rm shock}^{3}}{(p-2)(E_{0}^{2-p})} \quad \rm ergs^{-1} cm^{-2} s^{-1}
\end{equation}
Because we are only considering CRs for which $\mathcal{R}_{CR} < 1$, we now define $A_{p}$ as the normalization of the CR spectrum for energies only above $E_{\rm min}$. Then 

\begin{equation}
\label{fraction}
\frac{A_{\rm p}}{K_{\rm p}} = \Big(\frac{E_{\rm min}^{2-p}}{E_{0}^{2-p}} \Big)  
\end{equation}
This represents the fraction of CR luminosity in just the CRs that can diffuse back to the galaxy. The bottom panel of Figure \ref{CRProduced_contour} uses this fraction to obtain the luminosity of inward diffusing CRs for each solution to our wind model. These estimates assume $a = 0.4$ and the spectral slope $p = 2.2$. The fractions given by Equation (\ref{fraction}) are typically on the order of $10^{-1} - 10^{-3}$, resulting in luminosities of $10^{37} - 10^{39}$ ergs/s.

Cosmic ray interactions with circumgalactic gas can also produce gamma rays and neutrinos. \cite{2013ApJ...763...21F} estimate that these interactions in the Milky Way as well as other galaxies can contribute significantly to the observed isotropic gamma-ray background at energies above 1 GeV. Given the possibly large CR luminosities of termination shock accelerated CRs, it is worth calculating the gamma-ray and neutrino luminosities due to shock accelerated CRs in our wind models. A rigorous treatment of this will be the subject of future work.

\subsection{Estimates for the Milky Way Wind}
Equations (\ref{usefuldEdt}) and (\ref{reynolds}) can be applied to any wind model to estimate the cosmic ray production at the wind termination shock. Here, we do a short analysis of CR production due to the Milky Way wind using the results of \cite{2010ApJ...711...13E}. This paper fits the hybrid cosmic ray - thermally driven model of \cite{2008ApJ...674..258E} to the observed radio synchrotron and X-ray emission of the Milky Way wind to obtain the important wind parameters at various heights above the galactic disk. It should be noted that the \cite{2008ApJ...674..258E} model is not spherically symmetric. By construction, the wind flows through vertical tubes near the disk; however, these tubes open up at large radii, possibly giving the wind a more spherically symmetric profile far from the disk. The best fit parameters lead to $V_{\rm wind} \approx 600$ km/s and $n \approx 10^{-3} \rm cm^{-3}$ at $r = 10$ kpc. Assuming the wind evolves adiabatically in absence of radiative losses that are negligible at such low densities, we estimate that $R_{\rm shock} \approx 260$ kpc for a surrounding pressure of $P_{\rm IGM} = 10^{-14}$ ergs $\rm cm^{-3}$. Using Equation (\ref{dEdtEquipartition}), we find that for protons, $dE/dt = 6 \times 10^{-2} \rm GeV/yr$. For an acceleration time of 100 Myrs, protons can then be accelerated to $E_{\rm max} = 6 \times 10^{15} \rm eV$, not very high above the knee. A longer acceleration time would of course increase this energy, but with such a low wind velocity, one would expect that accelerating protons all the way to the ankle is virtually impossible. For a diffusion coefficient of $D_{0} = 5$ and $a = 0.4$, the minimum energy for diffusion-dominated CRs is $E_{\rm min} \approx 2.8 \times 10^{16}$ eV, greater than the maximum energy, meaning all CRs accelerated by the Milky Way wind termination shock will likely be advected away. Instead using $a = 0.5$, protons of energy $6 \times 10^{15}$ eV would have a good chance of diffusing back to the galaxy, as the cosmic ray Reynolds number we defined would only be about $\mathcal{R}_{CR} = 0.39$ for the shock velocity and radius derived above. In fact, the minimum energy of diffusion-dominated CRs in this case is $E_{\rm min} \approx 9.2 \times 10^{14}$ eV. 

Using $\dot{M} \sim 2 M_{\odot}/yr$, this Milky Way wind termination shock would optimistically produce a CR luminosity of $\eta \dot{M} V_{shock}^{2} \approx 4.5 \times 10^{40}$ ergs/s, including advection-dominated CRs. Assuming $p = 2.2$ for the CR spectrum and using the minimum and maximum CR energies estimated above for $a = 0.5$, the fraction of CR luminosity in inward diffusing CRs, given by Equation (\ref{fraction}), is $\approx 0.02$. Therefore, the luminosity of inward diffusing CRs is $L_{CR}^{\rm shock} \approx 9 \times 10^{38}$ ergs/s. This is only a few percent of the total CR luminosity in the Milky Way, which is $L_{CR}^{MW} \approx 10^{41}$ ergs/s. $L_{CR}^{\rm shock}$ only accounts for CRs with energies greater than $E_{\rm min} \approx 9.2 \times 10^{14}$ eV though.  \cite{2005neeu.conf....3G} estimates that the total luminosity of CRs with energies in the shin is roughly $2 \times 10^{39}$ ergs/s. Therefore, our estimate for $L_{CR}^{\rm shock}$ is more significant when considering just shin CRs; however, one should note that a more realistic estimate would be a fraction of this due to the Milky Way outflow being biconical instead of spherically symmetric. $L_{CR}^{\rm shock}$ is also very close to estimates of \cite{2014PhRvD..89j3003T} of the CR luminosity due to outflowing CRs in the galactic halo. They estimate $L_{CR}^{\rm halo} \approx 10^{39}$ ergs/s. Therefore, we conclude that our simple treatment of CR production leads to a possibly significant influx of shin CRs accelerated by the Milky Way wind termination shock.

\section{Conclusions}
\label{conclusions}
Suggestions that galactic wind termination shocks may accelerate shin CRs are generally well-motivated. CRs at these energies have gyroradii less than the typical distance to the termination shock, which is of order 100 kpc from the galaxy; prospects that this mechanism may explain the observed distribution of CRs, taking into account CR abundances, are good, though acceleration from Wolf-Rayet stars may give even better fits to the observed composition (\cite{2016arXiv160503111T}); and as in SNR shock fronts, magnetic field amplification could play a role in setting the necessary conditions on either side of the shock for efficient diffusive shock acceleration, although saturation of these amplification mechanisms is still not well understood. Advances in understanding of galactic outflows, including mounting evidence that winds are not in steady state for timescales of order a Gyr but instead follow timescales similar to those of short, intense episodes of star formation, points to the acceleration time of winds as a possible limiting factor for CR acceleration. Specifically, acceleration times on the order of 100 Myrs, which is also more in-line with the timescale of CR driven turbulent magnetic field amplification, necessitate wind velocities on the order of 2000 km/s to achieve $10^{17}$ eV CRs. This depends on the pressure of the surrounding IGM, which is uncertain. 

In our simplified, spherically symmetric picture of outflows, low mass-loaded winds typically have the highest velocities due to less-efficient radiative losses (BuZD16), and the velocity is fairly well set by the central sound speed, especially at high velocities. If conditions are correct, however, namely that wind velocities are greater than $\approx 700$ km/s for an IGM pressure of $10^{-14}$ ergs $ \rm cm^{-3}$, then CRs of order $10^{15} - 10^{16}$ eV may be accelerated by termination shocks over a reasonable time of 100 Myrs. Table \ref{windConditionsTable} in the Appendix shows model-independent wind velocities required to achieve various CR energies for varying acceleration times, as well as central temperature requirements specific to our model. We caution that an acceleration time of 1 Gyr is included for completeness; however, it is unlikely that strong outflows last for that long. 

Shin CRs accelerated by termination shocks might also be quite luminous, as shown in Figure \ref{CRProduced_contour} where $10^{42} - 10^{43}$ ergs/s, i.e. 1 - 10 \% of the Milky Way luminosity, is not out of the question. The luminosity of CRs that can diffuse back to the galaxy is a few orders of magnitude lower, as only a small fraction of CRs can overcome advection with the outflow. For the Milky Way, this luminosity of $\approx 9 \times 10^{38}$ ergs/s is possibly significant as we estimate it to be on the order of the total luminosity of observed shin CRs, but it is still orders of magnitude below the total observed CR luminosity of the Milky Way, which is $\sim 10^{41}$ ergs/s. 

Although wind velocities greater than 2000 km/s are theoretically possible in our model of thermally driven outflows, these values are quite high compared to those from observed stellar driven outflows. A few outflows on this order have been observed, however. X-ray fits suggest that the outflow of M82 has high terminal wind speeds of $\approx 1400 - 2200$ km/s (\cite{2009ApJ...697.2030S}). Similarly, \cite{2012ApJ...755L..26D} find outflows from compact, massive galaxies with velocities $> 1000$ km/s without the need for AGN feedback to explain the observations. 

The high thermal energy budget required to achieve 2000 km/s outflows suggests that galactic wind termination shocks, as presented here, may not be able to accelerate CRs past $ \approx 10^{17}$ eV. Interestingly, recent observations by IceTop have found another ``bump" in the CR energy spectrum near $10^{17}$ eV (\cite{2013PhRvD..88d2004A}) coinciding also with a change in CR composition. This is possibly suggestive of a change in acceleration mechanism. We can boldly speculate that this corresponds to the energy cutoff of CRs accelerated by termination shocks, but more work will need to be done before such claims can be made. 

Given constraints on the acceleration time and the high requirements for wind velocity, alternative, non-steady-state methods may be better-suited for explaining CRs at the high energy end of the shin. Specifically, \cite{2012A&A...540A..77D} suggest a pumping mechanism where time-varying galactic outflows pump energy into shocks that can then accelerate CRs. As opposed to a steady state model, this pumping mechanism could follow periodic bursts of star formation, possibly sustaining shocks for longer times. It is also attractive that these shocks could occur closer to the disk than a termination shock, making it more possible that the accelerated CRs can diffuse back to the galaxy. Another suggestion is that CRs could be re-accelerated by spiral shocks in the galaxy (\cite{2004A&A...417..807V}). These shocks can occur within 60 - 100 kpc of the galactic disk, which is favorable for diffusion back to the galactic disk. Assuming the shocks are long-lasting, the maximum energy of re-accelerated CRs can be determined by the condition $D_{B} \approx V_{\rm shock}R_{\rm shock}$ where $D_{B} = V_{\rm shock}r_{g}/3$ is the Bohm diffusion coefficient for particles with gyroradius $r_{g}$. Assuming an Archimedean spiral magnetic field at large distances, the estimated maximum CR energy is well beyond the knee; however, the lifetime of spiral shocks seems unknown. 

It should be stressed that the model-independent analysis in this paper is not limited to studies of just galactic wind termination shocks. The shocks created by the expanding ``cocoons" around AGN present very similar setups to that of galactic wind termination shocks. Given the massive amounts of energy in AGN driven outflows, these shocks could satisfy the high velocity requirements required for CR acceleration to ultra-high energies beyond the ankle. \cite{2016arXiv161107616W} estimate that non-relativistic quasar outflows can accelerate protons to $\approx 10^{20}$ eV using similar assumptions as in this paper. Further analysis of shock conditions and CR acceleration from AGN driven outflows is left to future work.

\section{Acknowledgements}
This material is based upon work supported by the National Science Foundation Graduate Research Fellowship Program under Grant No. DGE-1256259. Any opinions, findings, and conclusions or recommendations expressed in this material are those of the author(s) and do not necessarily reflect the views of the National Science Foundation. Support was also provided by the Graduate School and the Office of the Vice Chancellor for Research and Graduate Education at the University of Wisconsin-Madison with funding from the Wisconsin Alumni Research Foundation. CB and EGZ also acknowledge support from the University of Wisconsin-Madison and NSF Grant No. AST-1616037. CC acknowledges support from the Hilldale Undergraduate Research Fellowship at University of Wisconsin-Madison. The authors would like to thank the anonymous referee for the useful comments and suggestions. The authors also thank Dr. Francis Halzen, Dr. Jay Gallagher, and Dr. Tova Yoast-Hull for helpful discussions, as well as Dr. Alex Lazarian, whose special topics class influenced this work. 

\centering
\bibliographystyle{apj}

\bibliography{termShockPaper}

\newpage
\appendix
\label{appendix}
\begin{table}[!h]
\centering
\label{windConditionsTable}
\begin{tabular}{| c | c | c | c | c |}
\hline
$t_{\rm acc} = 10$ Myrs & $E_{CR}$ & Minimum $V_{\rm wind}$ & Minimum $T_{0}$ \\
\hline
& $10^{15}$ eV & 774 km/s  & $1.5 \times 10^{7}$ K  \\
& $10^{16}$ eV & 2440 km/s  & $1.4 \times 10^{8}$ K  \\
& $10^{17}$ eV & 7740 km/s  & $1.2 \times 10^{9}$ K \\
& $10^{18}$ eV & 24400 km/s  & $1.1 \times 10^{10}$ K \\
\hline
$t_{\rm acc} = 100$ Myrs & $E_{CR}$ & Minimum $V_{\rm wind}$ & Minimum $T_{0}$  \\
\hline
& $10^{15}$ eV & 244 km/s & $1.7 \times 10^{6}$ K \\
& $10^{16}$ eV & 774 km/s  & $1.5 \times 10^{7}$  K  \\
& $10^{17}$ eV & 2440 km/s & $1.4 \times 10^{8}$ K  \\
& $10^{18}$ eV & 7740 km/s  & $1.2 \times 10^{9}$ K \\
\hline

$t_{\rm acc} = 1$ Gyr & $E_{CR}$ & Minimum $V_{\rm wind}$ & Minimum $T_{0}$  \\
\hline
& $10^{15}$ eV & 77 km/s & $1.9 \times 10^{5}$ K  \\
& $10^{16}$ eV & 244 km/s  & $1.7 \times 10^{6}$ K \\
& $10^{17}$ eV & 774 km/s  & $1.5 \times 10^{7}$ K  \\
& $10^{18}$ eV & 2440 km/s  & $1.4 \times 10^{8}$ K \\
\hline
\end{tabular}
\caption{Target CR energies and estimates of the minimum wind velocity and minimum central temperature required to achieve those energies. The minimum wind velocity is derived for $P_{\rm IGM} = 10^{-14}$ ergs $\rm cm^{-3}$ by assuming an equipartition magnetic field and using the maximum acceleration rate obtained by \cite{1983A&A...125..249L}, which assumes Bohm diffusion at the shock. This information is independent of our wind model. The minimum central temperature, $T_{0}$, is gathered from the best fit line of Figure \ref{T0_vs_ShockVel_vs_Beta} for a desired asymptotic velocity.}
\end{table}

\end{document}